\documentclass[twocolumn,notitlepage,superscriptaddress]{revtex4-1} 

\usepackage[colorlinks=true,urlcolor=blue,linkcolor=blue,citecolor=blue]{hyperref} 
\usepackage{latexsym}
\usepackage{graphicx}
\usepackage{subfigure} 
\usepackage{placeins}
\usepackage{amsmath}
\usepackage{color}
\usepackage{amsfonts}
\usepackage{mathrsfs}
\usepackage{simpler-wick}
\usepackage{accents}

\providecommand \doibase [0]{http://dx.doi.org/} 

\makeatletter
\def\wideubar{\underaccent{{\cc@style\underline{\mskip10mu}}}}
\def\Wideubar{\underaccent{{\cc@style\underline{\mskip8mu}}}}
\makeatother
\makeatletter
\def\widebar{\accentset{{\cc@style\underline{\mskip10mu}}}}
\def\Widebar{\accentset{{\cc@style\underline{\mskip8mu}}}}
\makeatother

\begin{document}

\title{ Coincidence detection probability of $(\gamma, 2e)$ photoemission measurement}

\author{Yuehua Su}
\email{suyh@ytu.edu.cn}
\affiliation{ Department of Physics, Yantai University, Yantai 264005, People's Republic of China }

\author{Kun Cao}
\affiliation{ Department of Physics, Yantai University, Yantai 264005, People's Republic of China }

\author{Chao Zhang}
\affiliation{ Department of Physics, Yantai University, Yantai 264005, People's Republic of China }

\begin{abstract}

In the study of strongly correlated electrons, one of the challenging core tasks is to develop the potential techniques for direct detection of the many-body correlations of strongly correlated electrons. The $(\gamma, 2e)$ photoemission technique has been developed to investigate the two-body correlations of the target correlated electrons. In this article, we will focus on this technique for the correlated electrons near the Fermi energy in condensed matter. The coincidence detection probability of the two emitted electrons in the $(\gamma, 2e)$ photoemission measurement is shown to be relevant to a two-body Bethe-Salpeter wave function, which describes the dynamical two-body correlations of the target correlated electrons near the Fermi energy. As the coincidence detection probability involves an electron-electron interaction matrix element, the arbitrary momentum and/or energy transfer due to this electron-electron interaction makes the $(\gamma, 2e)$ photoemission technique fail to reveal the inner-pair structures of the two-body Bethe-Salpeter wave function. However, the center-of-mass momentum and energy of the two-body Bethe-Salpeter wave function can be distinctly resolved. Thus, the $(\gamma, 2e)$ photoemission technique can provide the center-of-mass physics of the two-body correlations of the target correlated electrons. It will be one potential technique to study the center-of-mass physics of the Cooper pairs in superconductor.      

\end{abstract}


\maketitle


\section{Introduction} \label{sec1}


The most challenge in the field of strongly correlated electrons is to understand the many-body correlations and the relevant experimental phenomena \citep{PALeeRMP2006, StewartNFLRMP2001, FradkinARCMP2010, StewartFeSCRMP2011, ChenXHNAR2014}. One core task in this field is to develop the potential techniques for direct detection of the many-body correlations of strongly correlated electrons. Recently, some coincidence detection techniques, the coincidence angle-resolved photoemission spectroscopy (cARPES) \citep{SuZhang2020,DevereauxPRB2023,CaoSu2024-2} and the coincidence inelastic neutron scattering (cINS) \citep{SucINS2021,CaoSu2024-2}, have been proposed to detect directly the two-body correlations of strongly correlated electrons. In the cARPES measurement, two incident photons emit two photoelectrons, respectively. By coincidence detection of two emitted photoelectrons, the cARPES can detect the coincidence probability of two relevant photoelectric processes. Thus, the cARPES can be developed to study the two-body correlations of the target correlated electrons in particle-particle channel. Similarly, by coincidence detection of two scattered neutrons, the cINS can detect the coincidence probability of two neutron-scattering processes, which can be developed to investigate the two-spin correlations of the target spin system. There are two basic principles in the theory of the proposed coincidence detection techniques: (1) The intrinsic two-body correlations can be detected directly by two-body coincidence detection, and (2) the coincidence detection probability is determined by the second-order perturbations of the interaction between the target matter and the external probe field. These basic principles can be developed to study the many-body correlations of the target matter in other channels. 

In this article, we will focus on an early proposed coincidence detection technique, the $(\gamma, 2e)$ photoemission technique which sometimes is called double photoemission technique. Unlike the cARPES technique, one single incident photon emits two electrons from the target matter in the $(\gamma, 2e)$ photoemission measurement. The $(\gamma, 2e)$ photoemission technique has been developed to study the core and the Auger electrons \citep{HaakAugerPRL1978}, the atomic and the molecular electrons \citep{WeberNature2005,2005H2Science,2007H2Science}, the valence electrons \citep{FominykhPRL2002,ChiangPRL2017},  the surface electrons \citep{Aliaev2018167}, the Cooper pairs \citep{BerakdarSCPRL2003}, {\it etc.} For each coincidence detection in the $(\gamma, 2e)$ photoemission measurement, there are mainly two types of contributions from two sequential microscopic physical processes. In type-I photoemission, the two sequential microscopic physical processes involve: (i) One incident photon is absorbed which excites one electron of the target matter into a high-energy intermediate state, and (ii) the excited electron interacts with another electron by an electron-electron interaction which scatters these two electrons out from the target matter in the final state. In type-II photoemission, the sequential microscopic physical processes include: (i) One incident photon is absorbed which emits one photoelectron from the target matter with one hole created, and (ii) two electrons in the target matter interact with each other with one excited to annihilate the created hole and the other scattered out in the final state. The two emitted electrons are detected coincidently in the $(\gamma, 2e)$ photoemission measurement and the coincidence detection probability involves the two-body correlations of the target-matter electrons. The two types of photoemission processes are different in the intermediate states. While an electron is created in the intermediate state of the type-I photoemission, it is a hole created in the intermediate state of the type-II photoemission. It should be noted that the Auger electron emission \citep{HaakAugerPRL1978} can also occur in type-II photoemission if the electrons we considered include the atomic core electrons. Therefore, the type-II photoemission can be called Auger-type photoemission.   

The theories for the $(\gamma, 2e)$ photoemission technique to study the correlated electrons near the Fermi energy in condensed matter have been developed recently \citep{BerakdarPRB1998, Fominykh2000665, BerakdarPRB2015}. In reference \citep{BerakdarPRB1998}, the author utilized the Fermi's golden rule to describe the single-photon absorption process and then introduced the time-reversed evolution technique to define the final state. With this theoretical treatment, the two sequential physical processes in the $(\gamma, 2e)$ photoemission measurement can be included in the coincidence detection probability. The references \citep{Fominykh2000665,BerakdarPRB2015} involve a similar idea in the definition of the coincidence current, where the initial and the final states are introduced with perturbation electron-electron interaction. These theories are too complex in their description of the dynamical properties of the two-body correlations of the target correlated electrons. Moreover, as we will show in this article, the coincidence detection probability is determined in principle by the second-order perturbations of the electron-photon and the electron-electron interactions. Therefore, a well-defined theory for the $(\gamma, 2e)$ photoemission measurement should be a second-order perturbation theory. 

In this article, we will develop a second-order perturbation theory for the $(\gamma, 2e)$ photoemission measurement. We will focus on the correlated electrons near the Fermi energy in condensed matter. Physically, the coincidence detection probability in the $(\gamma, 2e)$ photoemission measurement is defined by the coincidence probability of the two sequential microscopic physical processes, (i) the photon absorption process and (ii) the electron-electron scattering process. Mathematically, the coincidence detection probability is determined by the second-order perturbations of the two relevant electron-photon and electron-electron interactions. The coincidence detection probability is shown to be relevant to a dynamical two-body Bethe-Salpeter wave function, which describes the dynamical two-body correlations of the target correlated electrons near the Fermi energy. As the coincidence detection probability involves an electron-electron interaction matrix element which has arbitrary momentum and/or energy transfer, the inner-pair structures of the two-body Bethe-Salpeter wave function can not be resolved. Meanwhile, as the center-of-mass momentum and energy of the two-body Bethe-Salpeter wave function can be resolved, the $(\gamma, 2e)$ photoemission technique can be introduced to study the center-of-mass physics of the two-body correlations of the target correlated electrons, such as the center-of-mass physics of the Cooper pairs in superconductor.            

Our article is arranged as below. In Sec. \ref{sec2}, we will develop a second-order perturbation theory for the $(\gamma, 2e)$ photoemission measurement. In Sec. \ref{sec3}, we will consider a simplified case where the electron-electron interaction in the electron-electron scattering process is an instantaneous Coulomb interaction. We will present a summary in Sec. \ref{sec4} with a simple discussion on the coincidence detection of the Cooper-pair physics by the $(\gamma, 2e)$ photoemission technique.

\section{Theory for $(\gamma, 2e)$ photoemission measurement } \label{sec2}

In this section, we will present a second-order perturbation theory for the $(\gamma, 2e)$ photoemission measurement for the correlated electrons near the Fermi energy in condensed matter. 

As we have discussed in Sec. \ref{sec1}, each coincidence detection in the $(\gamma, 2e)$ photoemission measurement involves two sequential microscopic physical processes, (i) the photon absorption process and (ii) the electron-electron scattering process. The relevant electron-photon interaction $V_1$ and the electron-electron interaction $V_{2,T}$ of these two microscopic physical processes are defined as following: 
\begin{equation}
V_1(t) = \sum_{\mathbf{k} \mathbf{q}\lambda\sigma} g_\lambda(\mathbf{k},\mathbf{q}) d^\dag_{\mathbf{k}+\mathbf{q} \sigma}(t) c_{\mathbf{k}\sigma} (t) [a_{\mathbf{q}\lambda}(t) + a^\dag_{-\mathbf{q}\lambda}(t) ] , \label{eqn2.1}
\end{equation} 
and 
\begin{eqnarray}
V_{2,T}(t_1,t_2) &=& \frac{1}{2} \sum_{\mathbf{k}_1 \mathbf{k}_2 \mathbf{q}_1 \sigma_1 \sigma_2} U(\mathbf{k}_1, \mathbf{k}_2, \mathbf{q}_1; t_1, t_2)  f^\dag_{\mathbf{k}_1+\mathbf{q}_1 \sigma_1}(t_1) \notag \\
&& \times e_{\mathbf{k}_1 \sigma_1} (t_1) d^\dag_{\mathbf{k}_2 -\mathbf{q}_1 \sigma_2}(t_2)  c_{\mathbf{k}_2 \sigma_2 }(t_2) . \label{eqn2.2}
\end{eqnarray}
Here $c_{\mathbf{k}\sigma} ( d_{\mathbf{k}\sigma}, e_{\mathbf{k}\sigma}, f_{\mathbf{k}\sigma} )$ and $c^\dag_{\mathbf{k}\sigma} ( d^\dag_{\mathbf{k}\sigma}, e^\dag_{\mathbf{k}\sigma}, f^\dag_{\mathbf{k}\sigma} )$ are the annihilation and creation operators of the electrons with momentum $\mathbf{k}$ and spin $\sigma$, $a_{\mathbf{q}\lambda}$ and $a^\dag_{\mathbf{q}\lambda}$ are the annihilation and creation operators of the photons with momentum $\mathbf{q}$ and polarization $\lambda$. For clarity and brevity, we have introduced the different symbols $c, d, e, f$ for the electrons in the possible different electron bands. The band-index dependence of the interaction matrix elements $g$ and $U$ and the band-index summation are not explicitly shown. The more general forms of the electron-photon and the electron-electron interactions in crystal matter have been presented in Appendix \ref{secA}. Here the electron-photon interaction $V_1$ only involves the linear coupling of the electron field and the electromagnetic vector potential $\mathbf{A}$, as it is one relevant interaction of the $(\gamma, 2e)$ photoemission technique. This linear electron-photon interaction also plays major role in the various photoemission techniques, such as the conventional angle-resolved photoemission spectroscopy (ARPES) \citep{ShenRMP2003} and the recently developed pump-probe time-resolved angle-resolved photoemission spectroscopy (TR-ARPES) \citep{FreericksTRARPESPRL2009}. The TR-ARPES is a powerful technique to study the non-equilibrium dynamics of the target matter. The quadratic electron-$\mathbf{A}^2$ coupling may play unusual role in such as the cavity quantum electrodynamics (QED) experiments, where a strong attractive pairing interaction can be induced between two electrons, leading to a photoinduced Cooper pairing instability \citep{JakschPairingCavityPRL2020}.        

\begin{figure}[ht]
\includegraphics[width=1.0\columnwidth]{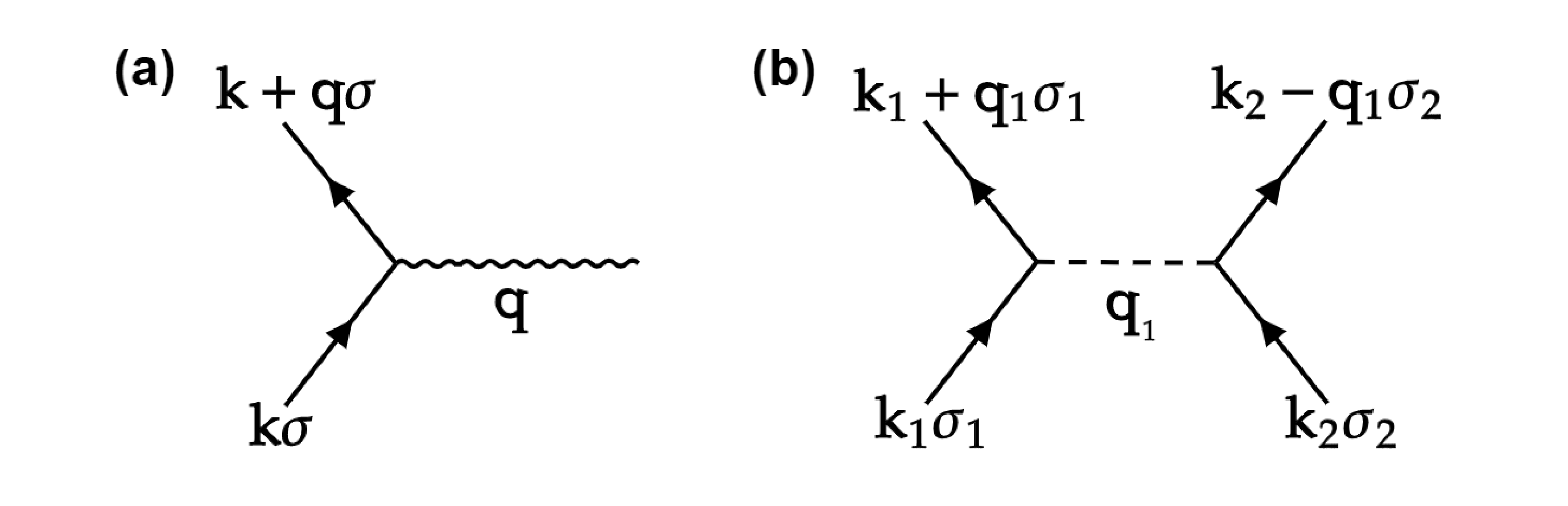} 
\caption{ Two relevant interactions in the $(\gamma, 2e)$ photoemission measurement, (a) the electron-photon interaction $V_1$, and (b) the electron-electron interaction $V_2$. }
\label{fig1}
\end{figure}

In the general cases, $V_{2,T}$ includes the Coulomb interaction and the phonon and/or other boson induced electron-electron interactions (such as from the charge or spin density fluctuations), the latter of which involve the time-retarded dynamics. In the below, we will assume that $U(\mathbf{k}_1, \mathbf{k}_2, \mathbf{q}_1; t_1, t_2)=U(\mathbf{k}_1, \mathbf{k}_2, \mathbf{q}_1; t_1- t_2)$ which follows energy conservation.  It should be remarked that for each coincidence detection with one given two sequential microscopic processes in the $(\gamma, 2e)$ photoemission measurement, only partial terms of $V_{2,T}$ are involved in the corresponding electron-electron scattering process. These partial terms of $V_{2,T}$, which will be named $V_2$ in the below, follow the constraints such as of the momentum, energy and spin conservation laws for the given initial and final states of each coincidence detection. Thus, $V_2$ is differently defined for different coincidence detection in the $(\gamma, 2e)$ photoemission measurement. The other terms of $V_{2,T}$ except $V_2$ contribute to the many-body correlations of the target correlated electrons. Therefore, the electron-electron interaction $V_{2,T}$ has a dual role.  The two relevant interactions for the $(\gamma, 2e)$ photoemission measurement can be schematically shown in Fig. \ref{fig1}.  

The Hamiltonian of the combined system of the $(\gamma, 2e)$ photoemission measurement is defined by 
\begin{equation}
H_T = H + V_1 + V_2 , \ H = H_e + H_p . \label{eqn2.3}
\end{equation}
Here $H_e$ is the Hamiltonian of the target correlated electrons, where all of the electron-electron interactions are involved except the ones in $V_2$. $H_p$ is the photon Hamiltonian defined by $H_p = \sum_{\mathbf{q}\lambda} \hbar \omega_\mathbf{q} (a^\dag_{\mathbf{q}\lambda} a_{\mathbf{q}\lambda}+\frac{1}{2})$. In the below, we will take the two interactions $V_1$ and $V_2$ as the perturbation interactions. We can define a time-evolution $S$-matrix operator as following:
\begin{equation}
S = T_t e^{-\frac{i}{\hbar} \left[ \int_{-\infty}^{+\infty} d t V_1(t) + \iint_{-\infty}^{+\infty} d t_1 d t_2 V_2(t_1, t_2) \right] } , \label{eqn2.4} 
\end{equation}
where $T_t$ is a time-ordering operator. 

Suppose the incident photon in the initial state has momentum $\mathbf{q}$ and polarization $\lambda$, and the two emitted electrons in the final state have momenta and spins $\mathbf{k}_1\sigma_1$ and $\mathbf{k}_2\sigma_2$, respectively. The coincidence detection probability of the two emitted electrons in the $(\gamma, 2e)$ photoemission measurement can be defined by 
\begin{equation}
\Gamma = \frac{1}{Z} \sum_{\alpha\beta} e^{-\beta E_\alpha} \big| \langle \Psi_\beta; \mathbf{k}_1\sigma_1,\mathbf{k}_2 \sigma_2 | S^{(2)} | \Psi_\alpha; \mathbf{q}\lambda \rangle \big|^2  , \label{eqn2.5} 
\end{equation}
where $| \Psi_\alpha \rangle$ and $| \Psi_\beta \rangle$ are the initial and the final states of the target correlated electrons in one coincidence detection, $E_\alpha$ is the corresponding eigenvalue of the eigenstate $| \Psi_\alpha \rangle$, $S^{(2)}$ is a second-order expansion of the $S$-matrix relevant to the $(\gamma, 2e)$ photoemission measurement and defined by, 
\begin{equation}
S^{(2)} = \frac{(-i)^2}{2!\hbar^2} \iiint_{-\infty}^{+\infty} d t_1 d t_2 d t_3 T_t [V_2(t_2, t_3) V_1(t_1)] . \label{eqn2.6}
\end{equation}
In Eq. (\ref{eqn2.5}), $P_{\alpha\beta}=\langle \Psi_\beta; \mathbf{k}_1\sigma_1,\mathbf{k}_2 \sigma_2 | S^{(2)} | \Psi_\alpha; \mathbf{q}\lambda \rangle$ defines the contribution from one coincidence detection to the coincidence detection probability $\Gamma$ and $\frac{1}{Z} e^{-\beta E_{\alpha}}$ defines the contribution statistics factor. It should be noted that here we have assumed an initial single-photon state $|\mathbf{q}\lambda \rangle$ in the definition of the coincidence detection probability $\Gamma$. This is just for discussion to be simple. In the realistic experiment, the incident photons can be in a macroscopic coherent state or other multiphoton state. In this case, the coincidence detection probability $\Gamma$ can be similarly defined with the new initial photon state. 

In order to describe the coincidence detection probability, following the theoretical treatment for cARPES \citep{SuZhang2020}, we introduce a two-body Bethe-Salpeter wave function \citep{SalpeterBethe1951, GellmanLowBS1951},
\begin{equation}
\Phi_{\alpha\beta}(\mathbf{k}_1 \sigma_1, t_1; \mathbf{k}_2 \sigma_2, t_2) = \langle \Psi_\beta | T_t c_{\mathbf{k}_1\sigma_1}(t_1) c_{\mathbf{k}_2\sigma_2} (t_2)  | \Psi_\alpha \rangle . \label{eqn2.7}
\end{equation}
It describes the physics of the target correlated electrons when two of the electrons are annihilated in time ordering. Thus, it describes the dynamical two-body correlations of the target correlated electrons in particle-particle channel. Introduce a center-of-mass time $t_c=\frac{1}{2}(t_1+t_2) $ and a relative time $t_r = t_1 - t_2$, the Bethe-Salpeter wave function can be reexpressed into  $\Phi_{\alpha\beta}(\mathbf{k}_1 \sigma_1, \mathbf{k}_2\sigma_2; t_c, t_r) = \Phi_{\alpha\beta}(\mathbf{k}_1 \sigma_1, t_1; \mathbf{k}_2 \sigma_2, t_2)$. The Fourier transformations of the two-body Bethe-Salpeter wave function are defined by
\begin{eqnarray}
&& \Phi_{\alpha\beta}(\mathbf{k}_1 \sigma_1, \mathbf{k}_2\sigma_2; t_c, t_r) \notag \\
&=&  \iint_{-\infty}^{+\infty} \frac{d \Omega d \omega}{(2\pi)^2} \Phi_{\alpha\beta}(\mathbf{k}_1 \sigma_1, \mathbf{k}_2\sigma_2; \Omega, \omega) e^{-i \Omega t_c - i \omega t_r} , \label{eqn2.8.1} \\
&& \Phi_{\alpha\beta}(\mathbf{k}_1 \sigma_1, \mathbf{k}_2\sigma_2; \Omega, \omega) \notag \\
&=&  \iint_{-\infty}^{+\infty} d t_c d t_r \Phi_{\alpha\beta}(\mathbf{k}_1 \sigma_1, \mathbf{k}_2\sigma_2; t_c, t_r)  e^{i \Omega t_c + i \omega t_r} . \label{eqn2.8.2}
\end{eqnarray}

\begin{figure}[ht]
\includegraphics[width=1.0\columnwidth]{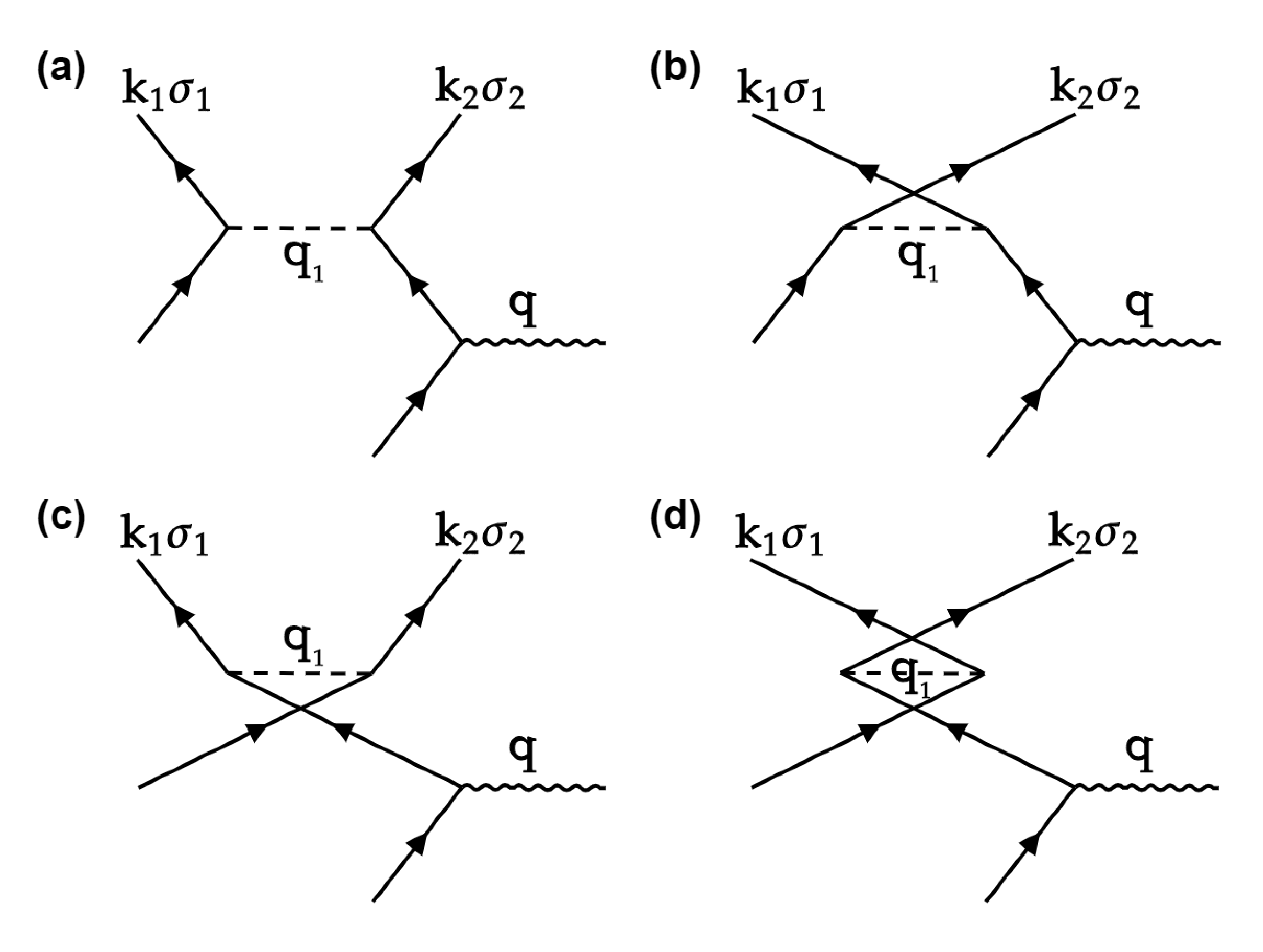} 
\caption{  Schematic illustration of the four contributions $P_{\alpha\beta,i}$ with $i=1,2,3,4$ to the coincidence detection probability $\Gamma_{\text I}$ in type-I photoemission. Here $i=1, 2, 3, 4$ correspond to (a), (b), (c) and (d), respectively. }
\label{fig2}
\end{figure}

Let us first consider the contributions from type-I photoemission processes in the $(\gamma, 2e)$ photoemission measurement. They are schematically shown in Fig. \ref{fig2}. In type-I photoemission, there are two sequential microscopic physical processes involved in each coincidence detection: (i) One incident photon is absorbed which excites one electron of the target matter into a high-energy intermediate state, and (ii) the excited electron interacts with another electron by the interaction $V_2$ which scatters these two electrons out from the target matter in the final state. The coincidence detection probability of the two emitted electrons from type-I photoemission can be shown to follow   
\begin{equation}
\Gamma_{\text I} = \frac{1}{Z}\sum_{\alpha\beta} e^{-\beta E_\alpha} \Big| \sum_{i=1}^4 P_{\alpha\beta,i}  \Big|^2 , \label{eqn2.9}
\end{equation}
where the four contributions $P_{\alpha\beta,i}, i =1, 2,3, 4$ to $\Gamma_{\text I}$ are schematically shown in Fig. \ref{fig2}. 

A detailed calculation of these four contributions is presented in Appendix \ref{secB}. It can be shown that $P_{\alpha\beta,1} = P_{\alpha\beta,1} (\mathbf{k}_1\sigma_1,\mathbf{k}_2\sigma_2)$ follows
\begin{eqnarray}
&&P_{\alpha\beta,1}(\mathbf{k}_1\sigma_1,\mathbf{k}_2\sigma_2) \notag \\
&=& \frac{-1}{4\hbar^2} \sum_{\mathbf{q}_1} g_\lambda(\mathbf{k}_2+\mathbf{q}_1-\mathbf{q},\mathbf{q}) U(\mathbf{k}_1-\mathbf{q}_1, \mathbf{k}_2+\mathbf{q}_1,\mathbf{q}_1;\omega_1) \notag \\
&& \times \Phi_{\alpha\beta}(\mathbf{k}_1-\mathbf{q}_1 \sigma_1,\mathbf{k}_2+\mathbf{q}_1 -\mathbf{q} \sigma_2; \Omega_c,\omega_r) , \label{eqn2.10}
\end{eqnarray}
where the frequencies $\omega_1$, $\Omega_c$ and $\omega_r$ are defined by 
\begin{eqnarray}
\omega_1 &=& \frac{1}{\hbar} [\varepsilon^{(I)}_{\mathbf{k}_2+\mathbf{q}_1} - \varepsilon^{(V)}_{\mathbf{k}_2} ] , \notag \\
\Omega_c &=& \frac{1}{\hbar} [\varepsilon^{(V)}_{\mathbf{k}_1} + \varepsilon^{(V)}_{\mathbf{k}_2} ] - \omega_{\mathbf{q}} , \label{eqn2.11} \\
\omega_r &=& \frac{1}{2\hbar} [\varepsilon^{(V)}_{\mathbf{k}_1} + \varepsilon^{(V)}_{\mathbf{k}_2} ] + \frac{1}{2}\omega_{\mathbf{q}} - \frac{1}{\hbar} \varepsilon^{(I)}_{\mathbf{k}_2+\mathbf{q}_1} . \notag 
\end{eqnarray}
Here $\omega_{\mathbf{q}}$ is the frequency of the incident photon, $\varepsilon_{\mathbf{k}}^{(I)}$ is the energy of the excited electron in the intermediate state after one photon absorption, and $\varepsilon_{\mathbf{k}}^{(V)}$ is the energy of the emitted electron. $U(\mathbf{k}_1, \mathbf{k}_2, \mathbf{q}_1; \omega_1)$ is the Fourier transformation of $U(\mathbf{k}_1, \mathbf{k}_2, \mathbf{q}; t_1-t_2)$ as defined by Eq. (\ref{eqnB.3}). It is clearly shown from Eq. (\ref{eqn2.10}) that $V_2$ involves only partial terms of the electron-electron interaction $V_{2,T}$ in one coincidence detection for $P_{\alpha\beta,1}(\mathbf{k}_1\sigma_1,\mathbf{k}_2\sigma_2)$, where the interaction matrix element $U(\mathbf{k}_1-\mathbf{q}_1, \mathbf{k}_2+\mathbf{q}_1,\mathbf{q}_1;\omega_1)$ has free momentum $\mathbf{q}_1$ but with all the other arguments fixed due to the momentum, energy and spin conservations. Similarly, $V_2$ involves only partial terms of $V_{2,T}$ in each coincidence detection for the other contributions, $P_{\alpha\beta,i}, i=2, \cdots, 8$. 
It should be noted that here we have made an assumption that the excited electron in the intermediate state can propagate freely in the target matter until it interacts with another electron by the interaction $V_2$. It can be a good approximation for the case where the excited electron in the intermediate state has a relatively long mean free path. Moreover, the sudden approximation \citep{ShenRMP2003} is also assumed for the two emitted electrons in the final state, which makes the momenta and energies of these two electrons conserved when they are tunnelling out from the target matter. In the above results, the work function has not been explicitly included for simplicity.  

The remaining three contributions $P_{\alpha\beta,2}, P_{\alpha\beta,3}$ and $P_{\alpha\beta,4}$ to $\Gamma_{\text I}$ follow the following equations:
\begin{equation}
P_{\alpha\beta,2}(\mathbf{k}_1\sigma_1,\mathbf{k}_2\sigma_2) = - P_{\alpha\beta,1}(\mathbf{k}_2\sigma_2,\mathbf{k}_1\sigma_1), \label{eqn2.12}
\end{equation}
and 
\begin{eqnarray}
&&P_{\alpha\beta,3}(\mathbf{k}_1\sigma_1,\mathbf{k}_2\sigma_2) = P_{\alpha\beta,2}(\mathbf{k}_1\sigma_1,\mathbf{k}_2\sigma_2), \notag \\
&&P_{\alpha\beta,4}(\mathbf{k}_1\sigma_1,\mathbf{k}_2\sigma_2) = P_{\alpha\beta,1}(\mathbf{k}_1\sigma_1,\mathbf{k}_2\sigma_2).  \label{eqn2.13}
\end{eqnarray}
Therefore, the coincidence detection probability of the two emitted electrons in type-I photoemission can be described by 
\begin{equation}
\Gamma_{\text I} = \frac{4}{Z}\sum_{\alpha\beta} e^{-\beta E_\alpha} \big| P_{\alpha\beta,1}(\mathbf{k}_1\sigma_1,\mathbf{k}_2\sigma_2) - P_{\alpha\beta,1}(\mathbf{k}_2\sigma_2,\mathbf{k}_1\sigma_1)  \big|^2 . \label{eqn2.14}
\end{equation}

\begin{figure}[ht]
\includegraphics[width=1.0\columnwidth]{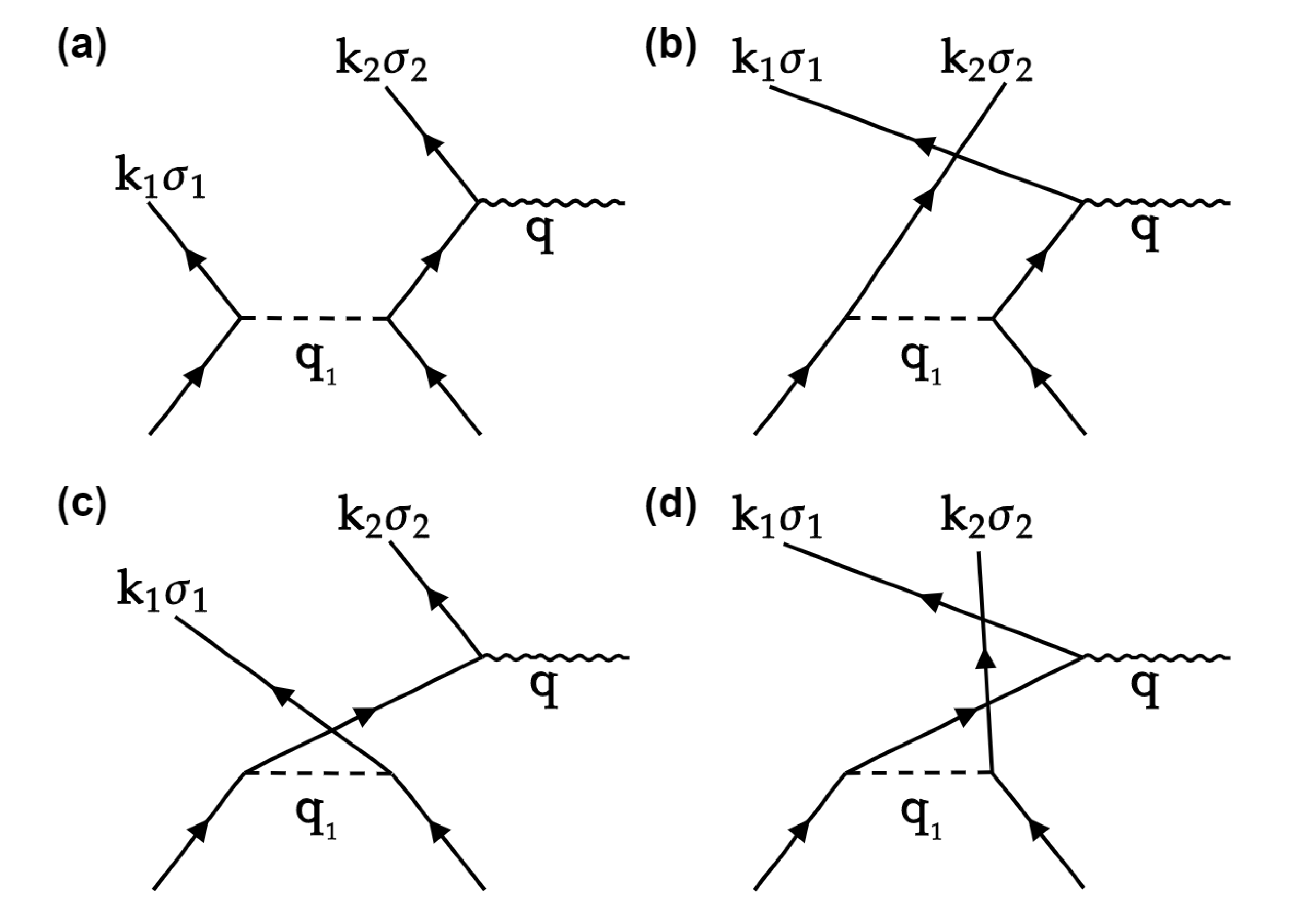} 
\caption{ Schematic illustration of the four contributions $P_{\alpha\beta,i}, i=5, 6, 7, 8$ to the coincidence detection probability $\Gamma_{\text{II}}$ in type-II photoemission. Here $i=5, 6, 7, 8$ correspond to (a), (b), (c) and (d), respectively. }
\label{fig3}
\end{figure}

Now let us study the contributions from type-II photoemission processes. There are four different microscopic contributions as schematically shown in Fig. \ref{fig3}. Each contribution involves two sequential microscopic physical processes: (i) One incident photon is absorbed which emits one photoelectron from the target matter with one hole created, and (ii) two electrons in the target matter interact with each other by the interaction $V_2$ with one excited to annihilate the created hole and the other scattered out in the final state. 

Following a similar calculation to type-I photoemission, we can show that (more details presented in Appendix \ref{secC})
\begin{eqnarray}
&&P_{\alpha\beta,6}(\mathbf{k}_1\sigma_1,\mathbf{k}_2\sigma_2) = - P_{\alpha\beta,5}(\mathbf{k}_2\sigma_2,\mathbf{k}_1\sigma_1), \notag \\
&& P_{\alpha\beta,8}(\mathbf{k}_1\sigma_1,\mathbf{k}_2\sigma_2) = - P_{\alpha\beta,7}(\mathbf{k}_2\sigma_2,\mathbf{k}_1\sigma_1). \label{eqn2.16}
\end{eqnarray}
Here $P_{\alpha\beta,5}(\mathbf{k}_1\sigma_1,\mathbf{k}_2\sigma_2)$ and $P_{\alpha\beta,7}(\mathbf{k}_1\sigma_1,\mathbf{k}_2\sigma_2)$ are given in the below: 
\begin{eqnarray}
&& P_{\alpha\beta,5}(\mathbf{k}_1\sigma_1,\mathbf{k}_2\sigma_2) \notag \\
&=& \frac{1}{4\hbar^2} \sum_{\mathbf{q}_1}\int d\omega_1 g_\lambda(\mathbf{k}_2-\mathbf{q},\mathbf{q}) U(\widebar{\mathbf{k}}_1, \widebar{\mathbf{k}}_2,\mathbf{q}_1;\omega_1) \notag \\
&& \times \Phi_{\alpha\beta}(\widebar{\mathbf{k}}_1 \sigma_1,\widebar{\mathbf{k}}_2 \sigma_2; \Omega_c,\omega_r) \delta(\omega_{\mathbf{q}} - \varepsilon^{(V)}_{\mathbf{k}_2}/\hbar + \varepsilon^{(I)}_{\mathbf{k}_2 - \mathbf{q}}/\hbar ) , \notag \\
&&  \label{eqn2.17}
\end{eqnarray}
where $\widebar{\mathbf{k}}_1 \equiv  \mathbf{k}_1-\mathbf{q}_1$ and $\widebar{\mathbf{k}}_2 \equiv  \mathbf{k}_2+\mathbf{q}_1 -\mathbf{q}$, and the frequencies $\Omega_c$ and $\omega_r$ are defined as 
\begin{eqnarray}
\Omega_c &=& \frac{1}{\hbar} [\varepsilon^{(V)}_{\mathbf{k}_1} + \varepsilon^{(V)}_{\mathbf{k}_2} ] - \omega_{\mathbf{q}} , \notag \\
\omega_r &=& \frac{1}{2\hbar} [\varepsilon^{(V)}_{\mathbf{k}_1} - \varepsilon^{(V)}_{\mathbf{k}_2} ] + \frac{1}{2}\omega_{\mathbf{q}} - \omega_1 . \label{eqn2.18} 
\end{eqnarray}
\begin{eqnarray}
&&P_{\alpha\beta,7}(\mathbf{k}_1\sigma_1,\mathbf{k}_2\sigma_2) \notag \\
&=& \frac{-1}{4\hbar^2} \sum_{\mathbf{q}_1}\int d\omega_1 g_\lambda(\mathbf{k}_2-\mathbf{q},\mathbf{q}) U(\widebar{\mathbf{k}}_2, \widebar{\mathbf{k}}_1,\mathbf{q}_1;\omega_1) \notag \\
&& \times \Phi_{\alpha\beta}(\widebar{\mathbf{k}}_2 \sigma_2,\widebar{\mathbf{k}}_1 \sigma_1; \Omega_c,\omega_r) \delta(\omega_{\mathbf{q}} - \varepsilon^{(V)}_{\mathbf{k}_2}/\hbar + \varepsilon^{(I)}_{\mathbf{k}_2 - \mathbf{q}}/\hbar ) , \notag \\
&& \label{eqn2.19}
\end{eqnarray}
where $\widebar{\mathbf{k}}_1 \equiv \mathbf{k}_1+\mathbf{q}_1  $ and $\widebar{\mathbf{k}}_2 \equiv \mathbf{k}_2-\mathbf{q}_1-\mathbf{q}$, and the frequencies $\Omega_c$ and $\omega_r$ are defined by 
\begin{eqnarray}
\Omega_c &=& \frac{1}{\hbar} [\varepsilon^{(V)}_{\mathbf{k}_1} + \varepsilon^{(V)}_{\mathbf{k}_2} ] - \omega_{\mathbf{q}} , \notag \\
\omega_r &=& \frac{1}{2\hbar} [\varepsilon^{(V)}_{\mathbf{k}_2} - \varepsilon^{(V)}_{\mathbf{k}_1} ] - \frac{1}{2}\omega_{\mathbf{q}} - \omega_1 . \label{eqn2.20} 
\end{eqnarray}
The coincidence detection probability from type-II phtoemission can be described by 
\begin{equation}
\Gamma_{\text{II}} = \frac{1}{Z}\sum_{\alpha\beta} e^{-\beta E_\alpha} \Big|\sum_{i=5}^{8} P_{\alpha\beta,i} \Big|^2 . \label{eqn2.21}
\end{equation}
In the derivation of the above results for type-II photoemission, we have also made an assumption that the excited electron can propagate freely before it annihilates the created hole. Meanwhile, the sudden approximation \citep{ShenRMP2003} is also assumed for the two emitted electrons in type-II photoemission.   

The total coincidence detection probability $\Gamma$ in the $(\gamma, 2e)$ photoemission measurement is defined by 
\begin{equation}
\Gamma = \Gamma_{\text I} + \Gamma_{\text{II}}, \label{eqn2.22}
\end{equation}
where $\Gamma_{\text I}$ and $\Gamma_{\text{II}}$ are given by Eqs. (\ref{eqn2.14}) and (\ref{eqn2.21}), respectively. Here we have ignored the quantum interference effects between the type-I and the type-II photoemission processes as they have different nondegenerate electron and hole intermediate states, respectively. The different energy conservation constraints of the two types of photoemission processes can not be satisfied simultaneously, which makes their quantum interference contribution zero. 
It should be noted that all of the eight contributions, $P_{\alpha\beta,i}, i=1, \cdots, 8$, to the total coincidence detection probability $\Gamma$ involve an electron-electron interaction matrix element $U(\mathbf{k}_1, \mathbf{k}_2, \mathbf{q}_1; \omega_1)$. This interaction matrix element involves the inter-electron momentum-$\mathbf{q}_1$ transfer between the two interacting electrons in the electron-electron scattering process. It leads to a momentum-$\mathbf{q}_1$ summation contribution to the coincidence detection probability. Moreover, the interaction matrix element also involves the inter-electron energy-$\omega_1$ transfer between the two interacting electrons. While the energy transfer in type-I photoemission is locked by the momentum transfer, the energy transfer in type-II photoemission is freely arbitrary. The arbitrary energy transfer leads to an additional frequency integral in the contributions to the coincidence detection probability from type-II photoemission. The arbitrary momentum and/or energy transfer makes the $(\gamma, 2e)$ photoemission technique fail to reveal the inner-pair spatial and dynamical structures of the two-body Bethe-Salpeter wave function. Thus, the $(\gamma, 2e)$ photoemission technique can not reveal the inner-pair physics of the two-body correlations of the target correlated electrons. It is not a good technique for direct detection of the pairing mechanism of the Cooper pairs in superconductor. 

Although the two interacting electrons have arbitrary momentum and/or energy transfer in the electron-electron scattering process, their center-of-mass momentum and energy are conserved. Thus, the center-of-mass momentum and energy of the two-body Bethe-Salpeter wave function can be resolved in the $(\gamma, 2e)$ photoemission measurement. Therefore, the coincidence detection probability of the $(\gamma, 2e)$ photoemission measurement can provide the center-of-mass physics of the two-body correlations of the target correlated electrons, with the electron-photon and the electron-electron interaction matrix elements $g$ and $U$ being the renormalization factors. The $(\gamma, 2e)$ photoemission technique can be introduced to study the center-of-mass physics of the two-body correlations of the target correlated electrons, \textit{e.g.},  the center-of-mass physics of the Cooper pairs in superconductor. It should be remarked that the proposed cARPES can provide both the inner-pair and the center-of-mass physics of the two-body correlations of the target correlated electrons \citep{SuZhang2020}.

\section{A simplified case with instantaneous Coulomb interaction} \label{sec3}

Let us consider a simplified case, where the electron-electron interaction $V_2$ is the instantaneous Coulomb interaction without time-retarded dynamics. This is a reasonable assumption for the case where the electron-electron scattering process has a much smaller time scale than that of the renormalization physical processes, such as the phonon, the charge or spin density fluctuation induced ones. In this case, the electron-electron scattering process occurs within a much short time so as the other charged particles, the charged ions and the other electrons, have not enough time to response this scattering process. Thus, the relevant electron-electron interaction $V_2$ can be approximated by an instantaneous interaction without time-retarded dynamics. This is reasonable for type-I photoemission where one of the two interacting electrons of the interaction $V_2$ is the excited electron in a high-energy intermediate state. For type-II photoemission, this is also reasonable if the created hole by the photon absorption has a much larger energy than the energy scales of the renormalization physical processes. For the other cases, the time-retarded renormalization effects should be included in the electron-electron interaction $V_2$.          

The instantaneous electron-electron interaction $V_2$ can be defined as partial terms of the following interaction $V_{2,T}$: 
\begin{eqnarray}
V_{2,T}(t) &=& \frac{1}{2} \sum_{\mathbf{k}_1 \mathbf{k}_2 \mathbf{q}_1 \sigma_1 \sigma_2}  U(\mathbf{q}_1)  f^\dag_{\mathbf{k}_1+\mathbf{q}_1 \sigma_1}(t) d^\dag_{\mathbf{k}_2 -\mathbf{q}_1 \sigma_2}(t) \notag \\
&& \times   c_{\mathbf{k}_2 \sigma_2 }(t) e_{\mathbf{k}_1 \sigma_1} (t) . \label{eqn3.1}
\end{eqnarray}
A general form of $U(\mathbf{q}_1)$ can be obtained with the same method provided in Appendix \ref{secA}. For the electrons in the approximate plane-wave states, the instantaneous electron-electron Coulomb interaction follows
\begin{equation}
U(\mathbf{q}_1) = \frac{e^2}{\varepsilon_0 V_D (|\mathbf{q}_1|^2 + q_\lambda^2 ) } , \label{eqn3.2}
\end{equation}
where $\varepsilon_0$ is the permittivity of vacuum and $V_D$ is the system volume. Here $q_\lambda = \frac{2\pi}{\lambda}$ where the screened length $\lambda$ comes from the spatial renormalization of the charged particles of the target matter. 

With a similar calculation to Appendix \ref{secB}, we can show that in the case with the instantaneous electron-electron interaction $V_2$, $P_{\alpha\beta,i}, i=1,2,3,4$ from type-I photoemission also follow Eq. (\ref{eqn2.12}) and (\ref{eqn2.13}), and
\begin{eqnarray}
&&P_{\alpha\beta,1}(\mathbf{k}_1\sigma_1,\mathbf{k}_2\sigma_2) \notag \\
&=& \frac{-1}{4\hbar^2} \sum_{\mathbf{q}_1} g_\lambda(\mathbf{k}_2+\mathbf{q}_1-\mathbf{q},\mathbf{q}) U(\mathbf{q}_1) \notag \\
&& \times \Phi_{\alpha\beta}(\mathbf{k}_1-\mathbf{q}_1 \sigma_1,\mathbf{k}_2+\mathbf{q}_1 -\mathbf{q} \sigma_2; \Omega_c,\omega_r) , \label{eqn3.3}
\end{eqnarray}
where the frequencies $\Omega_c$ and $\omega_r$ are defined by 
\begin{eqnarray}
\Omega_c &=& \frac{1}{\hbar} [\varepsilon^{(V)}_{\mathbf{k}_1} + \varepsilon^{(V)}_{\mathbf{k}_2} ] - \omega_{\mathbf{q}} , \notag \\
\omega_r &=& \frac{1}{2\hbar} [\varepsilon^{(V)}_{\mathbf{k}_1} + \varepsilon^{(V)}_{\mathbf{k}_2} ] + \frac{1}{2}\omega_{\mathbf{q}} - \frac{1}{\hbar} \varepsilon^{(I)}_{\mathbf{k}_2+\mathbf{q}_1} . \label{eqn2.26}
\end{eqnarray}
For type-II photoemission, we can show that in this case $P_{\alpha\beta,5}(\mathbf{k}_1\sigma_1,\mathbf{k}_2\sigma_2)$ and $P_{\alpha\beta,7}(\mathbf{k}_1\sigma_1,\mathbf{k}_2\sigma_2)$ follow as below: 
\begin{eqnarray}
&& P_{\alpha\beta,5}(\mathbf{k}_1\sigma_1,\mathbf{k}_2\sigma_2) \notag \\
&=& \frac{\pi}{2\hbar^2} \sum_{\mathbf{q}_1} g_\lambda(\mathbf{k}_2-\mathbf{q},\mathbf{q}) U(\mathbf{q}_1) \Phi_{\alpha\beta}(\widebar{\mathbf{k}}_1 \sigma_1,\widebar{\mathbf{k}}_2 \sigma_2; \Omega_c) \notag \\
&& \times  \delta(\omega_{\mathbf{q}} - \varepsilon^{(V)}_{\mathbf{k}_2}/\hbar + \varepsilon^{(I)}_{\mathbf{k}_2 - \mathbf{q}}/\hbar ) , \label{eqn3.4}
\end{eqnarray}
where $\widebar{\mathbf{k}}_1 \equiv  \mathbf{k}_1-\mathbf{q}_1$ and $\widebar{\mathbf{k}}_2 \equiv  \mathbf{k}_2+\mathbf{q}_1 -\mathbf{q}$, and the frequency $\Omega_c$ is defined as 
\begin{equation}
\Omega_c = \frac{1}{\hbar} [\varepsilon^{(V)}_{\mathbf{k}_1} + \varepsilon^{(V)}_{\mathbf{k}_2} ] - \omega_{\mathbf{q}}  , \label{eqn3.5} 
\end{equation}
and 
\begin{eqnarray}
&&P_{\alpha\beta,7}(\mathbf{k}_1\sigma_1,\mathbf{k}_2\sigma_2) \notag \\
&=& \frac{-\pi}{2\hbar^2} \sum_{\mathbf{q}_1} g_\lambda(\mathbf{k}_2-\mathbf{q},\mathbf{q}) U(\mathbf{q}_1) \Phi_{\alpha\beta}(\widebar{\mathbf{k}}_2 \sigma_2,\widebar{\mathbf{k}}_1 \sigma_1; \Omega_c) \notag \\
&& \times  \delta(\omega_{\mathbf{q}} - \varepsilon^{(V)}_{\mathbf{k}_2}/\hbar + \varepsilon^{(I)}_{\mathbf{k}_2 - \mathbf{q}}/\hbar ) ,  \label{eqn3.6}
\end{eqnarray}
where $\widebar{\mathbf{k}}_1 \equiv \mathbf{k}_1+\mathbf{q}_1  $ and $\widebar{\mathbf{k}}_2 \equiv \mathbf{k}_2-\mathbf{q}_1-\mathbf{q}$, and $\Omega_c$ is also defined by Eq. (\ref{eqn3.5}). Here $\Phi_{\alpha\beta}(\mathbf{k}_1 \sigma_1,\mathbf{k}_2 \sigma_2; \Omega_c) \equiv \Phi_{\alpha\beta}(\mathbf{k}_1 \sigma_1,\mathbf{k}_2 \sigma_2; \Omega_c, t_r=0)$. Moreover, it can be shown that the equations in Eq. (\ref{eqn2.16}) are also satisfied. The coincidence detection probability of the $(\gamma, 2e)$ photoemission measurement can also be similarly defined by Eqs. (\ref{eqn2.14}), (\ref{eqn2.21}) and (\ref{eqn2.22}). 

It is noted that in a recent article, the authors study the type-I photoemission contributions to the $(\gamma, 2e)$ photoemission measurement for different types of superconductors with a similarly defined instantaneous Coulomb interaction \citep{MorrPRB2022}. There is also a Coulomb-interaction relevant momentum summation in the coincidence detection probability, similar to that we have obtained in Eq. (\ref{eqn3.3}). One more remark is that the $(\gamma, 2e)$ photoemission technique can be introduced to study the Mott physics \citep{PALeeRMP2006} and the heavy fermion physics \citep{StewartNFLRMP2001} from the type-II photoemission contributions. Consider a Mott-Hubbard model with the Fermi energy in the higher-Hubbard band. Suppose the created hole in the intermediate state of the type-II photoemission process is in the lower-Hubbard band. The type-II photoemission probabilities, Eqs. (\ref{eqn2.17}), (\ref{eqn2.19}), (\ref{eqn3.4}) and (\ref{eqn3.6}), involve the two-body Bethe-Salpeter wave function of the electrons in the higher-Hubbard band as well as an energy $\delta$-function of the intermediate-state hole in the lower-Hubbard band. Therefore, the type-II photoemission can provide the correlated lower- and higher-Hubbard band Mott physics. Similarly, for the heavy fermion system when the created hole in the intermediate state is in the $f$-orbital, the type-II photoemission will provide the heavy fermion physics of the correlated $f$-orbital holes and the conduction-band electrons. Therefore, if the type-II photoemission contributions can be distinguished experimentally, the $(\gamma, 2e)$ photoemission technique will be a potential technique to study the Mott physics and the heavy fermions physics.

\section{Discussion and summary} \label{sec4}

In the above sections, we have developed a second-order perturbation theory for the $(\gamma, 2e)$ photoemission measurement for the correlated electrons near the Fermi energy in condensed matter. The coincidence detection probability of the two emitted electrons is relevant to a dynamical two-body Bethe-Salpeter wave function of the target correlated electrons. The dynamical two-body Bethe-Salpeter wave function has a general form \citep{SuZhang2020}:      
\begin{eqnarray}
\Phi_{\alpha\beta}\left(\mathbf{k}_1\sigma_1, \mathbf{k}_2\sigma_2; \Omega, \omega \right) &=& 2\pi \delta \left[\Omega + \left( E_{\beta} - E_{\alpha}\right)/\hbar \right] \notag \\
&& \times \phi_{\alpha\beta}\left(\mathbf{k}_1\sigma_1, \mathbf{k}_2\sigma_2; \omega \right) , \label{eqn4.1} 
\end{eqnarray}
where $\Phi_{\alpha\beta}\left(\mathbf{k}_1\sigma_1, \mathbf{k}_2\sigma_2; \omega \right)$ follows
\begin{eqnarray}
&&\Phi_{\alpha\beta}\left(\mathbf{k}_1\sigma_1, \mathbf{k}_2\sigma_2; \omega \right) \notag \\
 &=& \sum_{\gamma} \left[ \frac{ i \langle\Psi_{\beta} \vert c_{\mathbf{k}_1 \sigma_1} \vert \Psi_{\gamma} \rangle \langle \Psi_{\gamma} \vert  c_{\mathbf{k}_2 \sigma_2} \vert \Psi_{\alpha} \rangle} {\omega + i\delta^+ + (E_{\alpha} + E_{\beta} - 2 E_{\gamma} )/2\hbar} \right. \notag \\
&& \left. + \frac{ i \langle\Psi_{\beta} \vert c_{\mathbf{k}_2 \sigma_2} \vert \Psi_{\gamma} \rangle \langle \Psi_{\gamma} \vert  c_{\mathbf{k}_1 \sigma_1} \vert \Psi_{\alpha} \rangle} {\omega - i\delta^+ - (E_{\alpha} + E_{\beta} - 2 E_{\gamma} )/2\hbar} \right]. \label{eqn4.2}
\end{eqnarray}
It involves the following physics \citep{SuZhang2020}: (1) The center-of-mass dynamical physics  described by the $\delta$-function, $\delta \left[\Omega + \left( E_{\beta} - E_{\alpha}\right)/\hbar \right]$, which shows the energy transfer conservation in the center-of-mass channel. (2) The inner-pair dynamical physics described by $\Phi_{\alpha\beta}\left(\mathbf{k}_1\sigma_1, \mathbf{k}_2\sigma_2; \omega \right)$, which shows the propagator-like resonance structures peaked at $\omega = \pm (E_{\alpha} + E_{\beta} - 2 E_{\gamma} )/2\hbar$ with the weights defined by $\langle\Psi_{\beta} \vert c_{\mathbf{k}_1 \sigma_1} \vert \Psi_{\gamma} \rangle \langle \Psi_{\gamma} \vert  c_{\mathbf{k}_2 \sigma_2} \vert \Psi_{\alpha} \rangle$ and $\langle\Psi_{\beta} \vert c_{\mathbf{k}_2 \sigma_2} \vert \Psi_{\gamma} \rangle \langle \Psi_{\gamma} \vert  c_{\mathbf{k}_1 \sigma_1} \vert \Psi_{\alpha} \rangle$. 
As previously shown \citep{SuZhang2020}, the dynamical two-body Bethe-Salpeter wave function in a mean-field superconducting state has contributions from the macroscopic superconducting condensate and two Bogoliubov quasiparticles, the latter of which arise from the breaking of one Cooper pair. 

As we have discussed in Sec. \ref{sec2}, the $(\gamma, 2e)$ photoemission technique can not reveal the inner-pair structures of the two-body Bethe-Salpeter wave function due to the arbitrary momentum and/or energy transfer in the electron-electron scattering process in the $(\gamma, 2e)$ photoemission measurement. Since the center-of-mass momentum and energy of the dynamical two-body Bethe-Salpeter wave function can be resolved, the $(\gamma, 2e)$ photoemission technique can be used to study the center-of-mass physics of the two-body correlations of the target correlated electrons. As the Cooper pairs stem from the two-body correlations of the correlated electrons in superconductor, the $(\gamma, 2e)$ photoemission technique will be one potential technique to study the center-of-mass physics of the Cooper pairs in superconductor. 

Let us consider a spin-singlet superconductor. For the dynamical two-body Bethe-Salpeter wave function in $P_{\alpha\beta,1}$ of Eq. (\ref{eqn2.10}), $\Phi_{\alpha\beta}(\mathbf{k}_1-\mathbf{q}_1 \sigma_1,\mathbf{k}_2+\mathbf{q}_1 -\mathbf{q} \sigma_2; \Omega_c,\omega_r)$,  the focus on the Cooper pairs with the center-of-mass momentum $\mathbf{q}_c$ and frequency $\Omega_c$ leads to the following constraints:
\begin{eqnarray}
&&\mathbf{q}_c = \mathbf{k}_1 + \mathbf{k}_2 - \mathbf{q}, \notag \\
&&\Omega_c = \frac{1}{\hbar} [\varepsilon^{(V)}_{\mathbf{k}_1} + \varepsilon^{(V)}_{\mathbf{k}_2} ] - \omega_{\mathbf{q}},  \label{eqn4.3} \\
&&\sigma_1 = \uparrow (\downarrow), \sigma_2 = \downarrow (\uparrow) . \notag  
\end{eqnarray}  
Here $\mathbf{k}_1\sigma_1$ and $\mathbf{k}_2\sigma_2$ are the fixed momenta and spins of the two emitted electrons in the $(\gamma, 2e)$ photoemission measurement. The other dynamical two-body Bethe-Salpeter wave functions in $P_{\alpha,\beta, i}, i=2,\cdots,8$ also follow the same constraints. 

Physically, there are two types of collective excitations of the pairing gap field for the macroscopic superconducting condensate of a superconductor, the Goldstone phase modes and the Higgs amplitude modes. When the electromagnetic gauge fields are included, the Goldstone phase modes are modified into the plasmon modes by the so-called Anderson-Higgs mechanism \citep{AndersonPlasmon1958,NegeleOrland}. The plasmon modes are generally gapped with a large gap about $\hbar \omega_p \simeq 10 eV$ in most three-dimensional metallic superconductors. In the layered cuprate superconductors, the plasmon modes are highly anisotropic with strong out-of-plane momentum $q_{z}$ dependence  \citep{PlasmonsPRB1988,PlasmonsNature2018}. For the lowest-energy branch with $q_{z}=\frac{\pi}{d}$ where $d$ is the distance between neighbouring CuO$_2$ planes, the energy of the plasmon modes is highly suppressed with a similar feature to the layered electron gas. The Higgs amplitude modes are gapped with a gap value about two times of the Bogoliubov quasiparticle gap \citep{VarmaHiggsPRL1981,VarmaHiggsPRB1982}. In a two-band superconductor, the Leggett modes of the two macroscopic superconducting condensates may occur \citep{LeggettMode1966}. As the $(\gamma, 2e)$ photoemission technique is one potential technique to study the center-of-mass physics of the Cooper pairs in superconductor, it can be introduced to study these collective excitations of the macroscopic superconducting condensates.

In summary, we have developed a second-order perturbation theory for the $(\gamma, 2e)$ photoemission measurement for the correlated electrons near the Fermi energy in condensed matter. The coincidence detection probability of the two emitted electrons is relevant to a dynamical two-body Bethe-Salpeter wave function, which describes the dynamical two-body correlations of the target correlated electrons. Due to the arbitrary momentum and/or energy transfer in the electron-electron scattering process in the $(\gamma, 2e)$ photoemission measurement, only the center-of-mass momentum and energy of the dynamical two-body Bethe-Salpeter wave function can be resolved. Therefore, the $(\gamma, 2e)$ photoemission technique will be one potential technique to study the center-of-mass physics of the two-body correlations of the target correlated electrons, such as the center-of-mass physics of the Cooper pairs in superconductor.

\section*{ACKNOWLEDGMENTS}
We thank Prof. Jun Chang, Prof. Tao Li and and Prof. Shan Qiao for invaluable discussions. This work was supported by the National Natural Science Foundation of China (Grants No. 11774299 and No. 11874318) and the Natural Science Foundation of Shandong Province (Grant No. ZR2023MA015).

\begin{widetext}

\appendix

\section{Electron-photon and electron-electron interactions in crystal matter} \label{secA}

We will present the general second quantization forms of the electron-photon and the electron-electron interactions for the electrons in crystal matter. Let us first consider the electron-photon interaction, which origins from the gauge invariant Hamiltonian $H=\frac{1}{2m}(\mathbf{P}+ e \mathbf{A})^2$, where the charge of electron $q_e = -e$ has been included. Expand this Hamiltonian to linear-$\mathbf{A}$ terms and introduce the Coulomb gauge $\boldsymbol{\nabla}\cdot \mathbf{A}=0$, the electron-$\mathbf{A}$ interaction can be expressed as $V_1 = \frac{e}{m} \mathbf{A} \cdot \mathbf{P}$. Here we ignore the quadratic $\mathbf{A}^2$ term as it is little relevant to the $(\gamma, 2e)$ photoemission measurement. The second quantization of the electron-photon interaction can be obtained as following:
\begin{equation}
V_1 (t) = \sum_{\sigma}\int d \mathbf{r} \Psi_\sigma^{\dag}(\mathbf{r}, t) [\frac{e}{m} \mathbf{A} (\mathbf{r}, t)\cdot \mathbf{P}] \Psi_\sigma(\mathbf{r}, t), \label{eqnA.1}
\end{equation} 
where $\Psi_{\sigma}(\mathbf{r},t)$ and $\Psi^\dag_{\sigma}(\mathbf{r},t)$ are the electron annihilation and creation field operators, respectively. $\Psi_{\sigma}(\mathbf{r},t)$ can be expanded by the Bloch-band electron wave functions $\psi_n(\mathbf{k},\mathbf{r})$, 
\begin{equation}
\Psi_{\sigma}(\mathbf{r}, t) = \sum_{n\mathbf{k}} \psi_n(\mathbf{k},\mathbf{r}) c_{n\mathbf{k}\sigma} (t) , \label{eqnA.2}
\end{equation}
where $n$ is the band index number and $c_{n\mathbf{k}\sigma}$ is the electron annihilation operator. Here we have ignored the spin-orbit coupling effects on the Bloch-band electron wave functions and the momentum $\mathbf{k}$ is defined within the first Brillouin zone. From the Bloch theorem, we have 
\begin{equation}
\psi_n(\mathbf{k},\mathbf{r}) = \frac{1}{\sqrt{N\Omega_0}} \sum_{\mathbf{G}} a_n(\mathbf{k}+\mathbf{G}) e^{i(\mathbf{k}+\mathbf{G})\cdot \mathbf{r}} , \label{eqnA.3}
\end{equation} 
where $\mathbf{G}$ is the reciprocal lattice vector, $N$ is the lattice number, and $\Omega_0$ is the unit cell volume. Introduce the Fourier transformation of $\mathbf{A}$ field \citep{Bruus2002},  
\begin{equation}
\mathbf{A}(\mathbf{r}, t) = \sum_{\mathbf{q}\lambda}\sqrt{\frac{\hbar}{2\varepsilon_0 \omega_\mathbf{q} V_0}} \mathbf{e}_{\lambda}(\mathbf{q}) [a_{\mathbf{q}\lambda} (t) + a^\dag_{-\mathbf{q}\lambda} (t)] e^{i\mathbf{q}\cdot \mathbf{r}}, \label{eqnA.4} 
\end{equation}
where $\varepsilon_0$ is the permittivity of vacuum, $\omega_{\mathbf{q}}$ is the photon frequency, $V_0$ is the volume for $\mathbf{A}$ field to be enclosed, $\mathbf{e}_{\lambda}$ is the $\lambda$-th polarization unit vector, $a_{\mathbf{q}\lambda}$ and $a^\dag_{\mathbf{q}\lambda}$ are the photon annihilation and creation operators. From Eqs. (\ref{eqnA.2}), (\ref{eqnA.3}) and (\ref{eqnA.4}), the electron-photon interaction in the crystal matter can be expressed into the following form:
\begin{equation}
V_1 (t) = \sum_{n_1 n_2 \lambda}\sum_{\mathbf{k}\mathbf{q}\sigma} g_{n_1 n_2 \lambda}(\mathbf{k},\mathbf{q})  c^\dag_{n_1 \mathbf{k}+\mathbf{q}+\mathbf{G}\sigma} (t) c_{n_2 \mathbf{k}\sigma} (t)  [a_{\mathbf{q}\lambda} (t) + a^\dag_{-\mathbf{q}\lambda} (t)] , \label{eqnA.5} 
\end{equation}
where $g_{n_1 n_2 \lambda}(\mathbf{k},\mathbf{q}) $ is defined by 
\begin{equation}
g_{n_1 n_2 \lambda}(\mathbf{k},\mathbf{q}) =\sqrt{ \frac{e^2\hbar^3}{2 m^2 \varepsilon_0 \omega_\mathbf{q} V_0} }  \sum_{\mathbf{G}_l} [ \mathbf{e}_{\lambda}(\mathbf{q})\cdot (\mathbf{k}+\mathbf{G}_l) ] a^{\ast}_{n_1}(\mathbf{k}+\mathbf{q}+\mathbf{G}_l) a_{n_2}(\mathbf{k}+\mathbf{G}_l) . \label{eqnA.6}
\end{equation}
In Eq. (\ref{eqnA.5}), the reciprocal lattice vector $\mathbf{G}$ is defined so as for the given $\mathbf{k}$ and $\mathbf{q}$ with $\mathbf{k}$ in the first Brillouin zone, $\mathbf{k}+\mathbf{q}+\mathbf{G}$ is also in the first Brillouin zone.  

Now let us consider the electron-electron interaction which follows 
\begin{equation}
V_{2,T} (t) = \frac{1}{2}\sum_{\sigma_1 \sigma_2}\iint d\mathbf{r}_1 d\mathbf{r}_2 \Psi^\dag_{\sigma_1}(\mathbf{r}_1,t) \Psi^\dag_{\sigma_2}(\mathbf{r}_2, t) U(\mathbf{r}_1-\mathbf{r}_2) \Psi_{\sigma_2}(\mathbf{r}_2, t) \Psi_{\sigma_1}(\mathbf{r}_1, t) . \label{eqnA.7}
\end{equation}
With the electron fields expanded by the Bloch-band electron wave functions as Eq. (\ref{eqnA.2}), we can obtain the second quantization of the electron-electron interaction $V_{2,T}$ in the below form: 
\begin{equation}
V_{2,T} = \frac{1}{2}\sum_{n_1 n_2 n_3 n_4}\sum_{\mathbf{k}_1\mathbf{k}_2 \mathbf{q}\sigma_1\sigma_2} U_{n_1 n_2 n_3 n_4}(\mathbf{k}_1, \mathbf{k}_2,\mathbf{q}) c^\dag_{n_1 \mathbf{k}_1+\mathbf{q}+\mathbf{G}_1 \sigma_1} (t) c^\dag_{n_2 \mathbf{k}_2-\mathbf{q}+\mathbf{G}_2 \sigma_2} (t) c_{n_3 \mathbf{k}_2 \sigma_2} (t) c_{n_4 \mathbf{k}_1 \sigma_1} (t) , \label{eqnA.8} 
\end{equation}
where $ U_{n_1 n_2 n_3 n_4}(\mathbf{k}_1, \mathbf{k}_2,\mathbf{q})$ is defined by 
\begin{equation}
U_{n_1 n_2 n_3 n_4}(\mathbf{k}_1, \mathbf{k}_2,\mathbf{q}) = \sum_{\mathbf{G}_l \mathbf{G}_m} U(\mathbf{q}) a^{\ast}_{n_1}(\mathbf{k}_1+\mathbf{q}+\mathbf{G}_l) a^{\ast}_{n_2}(\mathbf{k}_2-\mathbf{q}+\mathbf{G}_m) a_{n_3}(\mathbf{k}_2+\mathbf{G}_m) a_{n_4}(\mathbf{k}_1+\mathbf{G}_l) . \label{eqnA.9}
\end{equation}
Here $\mathbf{G}_{1}$ and $\mathbf{G}_{2}$ in Eq. (\ref{eqnA.8}) are two reciprocal lattice vectors which make $\mathbf{k}_1+\mathbf{q}+\mathbf{G}_1$ and $\mathbf{k}_2-\mathbf{q}+\mathbf{G}_2$ to be in the first Brillouin zone, respectively. During the derivation, we have introduced the Fourier transformation of $U(\mathbf{r})$, $U(\mathbf{r})=\sum_{\mathbf{q}} U(\mathbf{q}) e^{i \mathbf{q}\cdot \mathbf{r}}$.

\section{Calculation of $P_{\alpha\beta,i}, i=1,2,3,4$ in type-I photoemission } \label{secB}

In this Appendix section, we will present a detailed study on $P_{\alpha\beta,i}, i=1,2,3,4$ in type-I photoemission. They are schematically shown in Fig. \ref{fig2}. 
Let us first consider $P_{\alpha\beta,1}$. It can be calculated as follows. 
\begin{eqnarray}
P_{\alpha\beta,1} &=& \frac{(-i)^2}{2! \hbar^2} \iiint_{-\infty}^{+\infty} d t_1 d t_2 d t_3 \sum_{\mathbf{k}^{\prime}_1 \mathbf{k}^{\prime}_2 \mathbf{q}^{\prime}_1 \sigma^{\prime}_1 \sigma^{\prime}_2} \sum_{\mathbf{k}^{\prime} \mathbf{q}^{\prime} \lambda^{\prime} \sigma^{\prime}} \frac{1}{2}	U(\mathbf{k}^{\prime}_1, \mathbf{k}^{\prime}_2, \mathbf{q}^{\prime}_1; t_2 - t_3) g_{\lambda^{\prime}}(\mathbf{k}^{\prime},\mathbf{q}^{\prime})  \notag \\
&& \times 
\wick{
\langle \Psi_\beta; \mathbf{k}_{1} \c1 \sigma_1 ,\mathbf{k}_2 \c2 \sigma_2 | T_t  \c1 f^\dag_{\mathbf{k}^{\prime}_1+\mathbf{q}^{\prime}_1 \sigma^{\prime}_1}(t_2) c_{\mathbf{k}^{\prime}_1 \sigma^{\prime}_1} (t_2) \c2 f^\dag_{\mathbf{k}^{\prime}_2 -\mathbf{q}^{\prime}_1 \sigma^{\prime}_2}(t_3) \c1 d_{\mathbf{k}^{\prime}_2 \sigma^{\prime}_2 }(t_3) \c1 d^\dag_{\mathbf{k}^{\prime}+\mathbf{q}^{\prime} \sigma^{\prime}}(t_1) c_{\mathbf{k}^{\prime}\sigma^{\prime}} (t_1) [ \c1 a_{ \mathbf{q}^{\prime}\lambda^{\prime}}(t_1) + a^\dag_{-\mathbf{q}^{\prime}\lambda^{\prime}} (t_1) ]   | \Psi_{\alpha}; \mathbf{q} \c1 \lambda \rangle
}
 \notag \\
&=& \frac{-1}{4 \hbar^2} \iiint_{-\infty}^{+\infty} d t_1 d t_2 d t_3 \sum_{\mathbf{k}^{\prime}_1 \mathbf{k}^{\prime}_2 \mathbf{q}^{\prime}_1 \sigma^{\prime}_1 \sigma^{\prime}_2} \sum_{\mathbf{k}^{\prime}  \sigma^{\prime}} U(\mathbf{k}^{\prime}_1, \mathbf{k}^{\prime}_2, \mathbf{q}^{\prime}_1; t_2 - t_3) g_{\lambda}(\mathbf{k}^{\prime},\mathbf{q})      e^{-i\omega_{\mathbf{q}} t_1 + i \varepsilon^{(V)}_{\mathbf{k}^{\prime}_1+\mathbf{q}^{\prime}_1} t_2/\hbar + i \varepsilon^{(V)}_{\mathbf{k}^{\prime}_2-\mathbf{q}^{\prime}_1} t_3/\hbar } \notag \\
&& \times 
\wick{
\langle \Psi_\beta | T_t  c_{\mathbf{k}^{\prime}_1 \sigma^{\prime}_1} (t_2) \c1 d_{\mathbf{k}^{\prime}_2 \sigma^{\prime}_2 }(t_3) \c1 d^\dag_{\mathbf{k}^{\prime}+\mathbf{q} \sigma^{\prime}}(t_1) c_{\mathbf{k}^{\prime}\sigma^{\prime}} (t_1)   | \Psi_{\alpha} \rangle
}
\delta_{\mathbf{k}_1, \mathbf{k}^\prime+\mathbf{q}^\prime_1}\delta_{\sigma_1 \sigma_1^\prime} \delta_{\mathbf{k}_2,\mathbf{k}^\prime_2-\mathbf{q}_1^\prime} \delta_{\sigma_2 \sigma_2^\prime} \notag \\
&=& \frac{-1}{4 \hbar^2} \iiint_{-\infty}^{+\infty} d t_1 d t_2 d t_3 \sum_{\mathbf{k}^{\prime}_1 \mathbf{k}^{\prime}_2 \mathbf{q}^{\prime}_1 \sigma^{\prime}_1 \sigma^{\prime}_2} \sum_{\mathbf{k}^{\prime} \sigma^{\prime}} U(\mathbf{k}^{\prime}_1, \mathbf{k}^{\prime}_2, \mathbf{q}^{\prime}_1; t_2 - t_3) g_{\lambda}(\mathbf{k}^{\prime},\mathbf{q})      e^{-i\omega_{\mathbf{q}} t_1 + i \varepsilon^{(V)}_{\mathbf{k}^{\prime}_1+\mathbf{q}^{\prime}_1} t_2/\hbar + i \varepsilon^{(V)}_{\mathbf{k}^{\prime}_2-\mathbf{q}^{\prime}_1} t_3/\hbar } \notag \\
&& \times 
\langle \Psi_\beta | T_t  c_{\mathbf{k}^{\prime}_1 \sigma^{\prime}_1} (t_2) c_{\mathbf{k}^{\prime}\sigma^{\prime}} (t_1)   | \Psi_{\alpha} \rangle
\delta_{\mathbf{k}_1, \mathbf{k}^\prime+\mathbf{q}^\prime_1}\delta_{\sigma_1 \sigma_1^\prime} \delta_{\mathbf{k}_2,\mathbf{k}^\prime_2-\mathbf{q}_1^\prime} \delta_{\sigma_2 \sigma_2^\prime} \delta_{\mathbf{k}_2^\prime,\mathbf{k}^\prime+\mathbf{q}} \delta_{\sigma_2^\prime \sigma^\prime} e^{-i\varepsilon^{(I)}_{\mathbf{k}_2^\prime}(t_3 - t_1)/\hbar } .  \label{eqnB.1} 
\end{eqnarray}
Here the contraction operations, unlike the common definition for the Green's functions in the many-body perturbation theory, describe the creation-and-annihilation relations. In the second step, the emitted electron from the target matter with momentum $\mathbf{k}$ and spin $\sigma$ is assumed to have energy $\varepsilon^{(V)}_{\mathbf{k}}$, and $a_{\mathbf{q}\lambda}(t)=a_{\mathbf{q}\lambda} e^{-i\omega_{\mathbf{q}} t}$ is used where $\omega_{\mathbf{q}}$ is the photon frequency. In the last step, we assume that the electron in the intermediate state excited by the incident photon can propagate freely in the target matter with energy $\varepsilon^{(I)}_{\mathbf{k}_2^\prime}$, thus we have   
\begin{equation}
\wick{
\c1 d_{\mathbf{k}^{\prime}_2 \sigma^{\prime}_2 }(t_3) \c1 d^\dag_{\mathbf{k}^{\prime}+\mathbf{q} \sigma^{\prime}}(t_1) 
} = \delta_{\mathbf{k}_2^\prime,\mathbf{k}^\prime+\mathbf{q}} \delta_{\sigma_2^\prime \sigma^\prime} e^{-i\varepsilon^{(I)}_{\mathbf{k}_2^\prime}(t_3 - t_1)/\hbar } . \label{eqnB.2}
\end{equation}
This is one approximate assumption in our theory for the coincidence detection probability in the $(\gamma, 2e)$ photoemission measurement. Let us introduce the Fourier transformation of $U$,
\begin{eqnarray}
U(\mathbf{k}^{\prime}_1, \mathbf{k}^{\prime}_2, \mathbf{q}^{\prime}_1; t_2 - t_3) = \frac{1}{2 \pi} \int d\omega_1 U(\mathbf{k}^{\prime}_1, \mathbf{k}^{\prime}_2, \mathbf{q}^{\prime}_1; \omega_1) e^{-i\omega_1 (t_2 - t_3)} . \label{eqnB.3}
\end{eqnarray}
Note that there is one symmetry for $U$: $U(\mathbf{k}^{\prime}_1, \mathbf{k}^{\prime}_2, \mathbf{q}^{\prime}_1; t_2-t_3) = U(\mathbf{k}^{\prime}_2, \mathbf{k}^{\prime}_1, -\mathbf{q}^{\prime}_1; t_3-t_2)$. With the two-body Bethe-Salpeter wave function defined in Eq. (\ref{eqn2.7}) and the Fourier transformations in Eqs. (\ref{eqn2.8.1}) and (\ref{eqn2.8.2}), we can show that $P_{\alpha\beta,1}$ follows Eq. (\ref{eqn2.10}). 

The definitions of $P_{\alpha\beta,2},P_{\alpha\beta,3}$ and $P_{\alpha\beta,4}$ are given as below:
\begin{eqnarray}
P_{\alpha\beta,2} &=& \frac{(-i)^2}{2! \hbar^2} \iiint_{-\infty}^{+\infty} d t_1 d t_2 d t_3 \sum_{\mathbf{k}^{\prime}_1 \mathbf{k}^{\prime}_2 \mathbf{q}^{\prime}_1 \sigma^{\prime}_1 \sigma^{\prime}_2} \sum_{\mathbf{k}^{\prime} \mathbf{q}^{\prime} \lambda^{\prime} \sigma^{\prime}} \frac{1}{2}	U(\mathbf{k}^{\prime}_1, \mathbf{k}^{\prime}_2, \mathbf{q}^{\prime}_1; t_2 - t_3) g_{\lambda^{\prime}}(\mathbf{k}^{\prime},\mathbf{q}^{\prime})  \notag \\
&& \times 
\wick{
\langle \Psi_\beta; \mathbf{k}_{1} \c2 \sigma_1 ,\mathbf{k}_2 \c1 \sigma_2 | T_t  \c1 f^\dag_{\mathbf{k}^{\prime}_1+\mathbf{q}^{\prime}_1 \sigma^{\prime}_1}(t_2) c_{\mathbf{k}^{\prime}_1 \sigma^{\prime}_1} (t_2) \c2 f^\dag_{\mathbf{k}^{\prime}_2 -\mathbf{q}^{\prime}_1 \sigma^{\prime}_2}(t_3) \c1 d_{\mathbf{k}^{\prime}_2 \sigma^{\prime}_2 }(t_3) \c1 d^\dag_{\mathbf{k}^{\prime}+\mathbf{q}^{\prime} \sigma^{\prime}}(t_1) c_{\mathbf{k}^{\prime}\sigma^{\prime}} (t_1) [ \c1 a_{ \mathbf{q}^{\prime}\lambda^{\prime}}(t_1) + a^\dag_{-\mathbf{q}^{\prime}\lambda^{\prime}} (t_1) ]   | \Psi_{\alpha}; \mathbf{q} \c1 \lambda \rangle
}
 \notag \\
P_{\alpha\beta,3} &=& \frac{(-i)^2}{2! \hbar^2} \iiint_{-\infty}^{+\infty} d t_1 d t_2 d t_3 \sum_{\mathbf{k}^{\prime}_1 \mathbf{k}^{\prime}_2 \mathbf{q}^{\prime}_1 \sigma^{\prime}_1 \sigma^{\prime}_2} \sum_{\mathbf{k}^{\prime} \mathbf{q}^{\prime} \lambda^{\prime} \sigma^{\prime}} \frac{1}{2}	U(\mathbf{k}^{\prime}_1, \mathbf{k}^{\prime}_2, \mathbf{q}^{\prime}_1; t_2 - t_3) g_{\lambda^{\prime}}(\mathbf{k}^{\prime},\mathbf{q}^{\prime})  \label{eqnB.4} \\
&& \times 
\wick{
\langle \Psi_\beta; \mathbf{k}_{1} \c1 \sigma_1 ,\mathbf{k}_2 \c2 \sigma_2 | T_t  \c1 f^\dag_{\mathbf{k}^{\prime}_1+\mathbf{q}^{\prime}_1 \sigma^{\prime}_1}(t_2) \c1 d_{\mathbf{k}^{\prime}_1 \sigma^{\prime}_1} (t_2) \c2 f^\dag_{\mathbf{k}^{\prime}_2 -\mathbf{q}^{\prime}_1 \sigma^{\prime}_2}(t_3) c_{\mathbf{k}^{\prime}_2 \sigma^{\prime}_2 }(t_3) \c1 d^\dag_{\mathbf{k}^{\prime}+\mathbf{q}^{\prime} \sigma^{\prime}}(t_1) c_{\mathbf{k}^{\prime}\sigma^{\prime}} (t_1) [ \c1 a_{ \mathbf{q}^{\prime}\lambda^{\prime}}(t_1) + a^\dag_{-\mathbf{q}^{\prime}\lambda^{\prime}} (t_1) ]   | \Psi_{\alpha}; \mathbf{q} \c1 \lambda \rangle
}
 \notag \\ 
 P_{\alpha\beta,4} &=& \frac{(-i)^2}{2! \hbar^2} \iiint_{-\infty}^{+\infty} d t_1 d t_2 d t_3 \sum_{\mathbf{k}^{\prime}_1 \mathbf{k}^{\prime}_2 \mathbf{q}^{\prime}_1 \sigma^{\prime}_1 \sigma^{\prime}_2} \sum_{\mathbf{k}^{\prime} \mathbf{q}^{\prime} \lambda^{\prime} \sigma^{\prime}} \frac{1}{2}	U(\mathbf{k}^{\prime}_1, \mathbf{k}^{\prime}_2, \mathbf{q}^{\prime}_1; t_2 - t_3) g_{\lambda^{\prime}}(\mathbf{k}^{\prime},\mathbf{q}^{\prime})  \notag \\
&& \times 
\wick{
\langle \Psi_\beta; \mathbf{k}_{1} \c2 \sigma_1 ,\mathbf{k}_2 \c1 \sigma_2 | T_t  \c1 f^\dag_{\mathbf{k}^{\prime}_1+\mathbf{q}^{\prime}_1 \sigma^{\prime}_1}(t_2) \c1 d_{\mathbf{k}^{\prime}_1 \sigma^{\prime}_1} (t_2) \c2 f^\dag_{\mathbf{k}^{\prime}_2 -\mathbf{q}^{\prime}_1 \sigma^{\prime}_2}(t_3) c_{\mathbf{k}^{\prime}_2 \sigma^{\prime}_2 }(t_3) \c1 d^\dag_{\mathbf{k}^{\prime}+\mathbf{q}^{\prime} \sigma^{\prime}}(t_1) c_{\mathbf{k}^{\prime}\sigma^{\prime}} (t_1) [ \c1 a_{ \mathbf{q}^{\prime}\lambda^{\prime}}(t_1) + a^\dag_{-\mathbf{q}^{\prime}\lambda^{\prime}} (t_1) ]   | \Psi_{\alpha}; \mathbf{q} \c1 \lambda \rangle
}
 \notag 
\end{eqnarray} 
With a similar calculation to $P_{\alpha\beta,1}$, we can show that $P_{\alpha\beta,i}, i=2,3,4$ follow Eqs. (\ref{eqn2.12}) and (\ref{eqn2.13}). 

\section{Calculation of $P_{\alpha\beta,i}, i=5,6,7,8$ in type-II photoemission} \label{secC}

In this Appendix section, we will calculate the contributions from type-II photoemission to the coincidence detection probability. They are described by $P_{\alpha\beta,i}, i=5,6,7,8$, which are schematically shown in Fig. \ref{fig3}. Mathematically, they are defined as below: 
\begin{eqnarray}
P_{\alpha\beta,5} &=& \frac{(-i)^2}{2! \hbar^2} \iiint_{-\infty}^{+\infty} d t_1 d t_2 d t_3 \sum_{\mathbf{k}^{\prime}_1 \mathbf{k}^{\prime}_2 \mathbf{q}^{\prime}_1 \sigma^{\prime}_1 \sigma^{\prime}_2} \sum_{\mathbf{k}^{\prime} \mathbf{q}^{\prime} \lambda^{\prime} \sigma^{\prime}} \frac{1}{2}	U(\mathbf{k}^{\prime}_1, \mathbf{k}^{\prime}_2, \mathbf{q}^{\prime}_1; t_2 - t_3) g_{\lambda^{\prime}}(\mathbf{k}^{\prime},\mathbf{q}^{\prime})  \notag \\
&& \times 
\wick{
\langle \Psi_\beta; \mathbf{k}_{1} \c1 \sigma_1 ,\mathbf{k}_2 \c2 \sigma_2 | T_t  \c1 f^\dag_{\mathbf{k}^{\prime}_1+\mathbf{q}^{\prime}_1 \sigma^{\prime}_1}(t_2) c_{\mathbf{k}^{\prime}_1 \sigma^{\prime}_1} (t_2) \c1 d^\dag_{\mathbf{k}^{\prime}_2 -\mathbf{q}^{\prime}_1 \sigma^{\prime}_2}(t_3) c_{\mathbf{k}^{\prime}_2 \sigma^{\prime}_2 }(t_3) \c2 f^\dag_{\mathbf{k}^{\prime}+\mathbf{q}^{\prime} \sigma^{\prime}}(t_1) \c1 d_{\mathbf{k}^{\prime}\sigma^{\prime}} (t_1) [ \c1 a_{ \mathbf{q}^{\prime}\lambda^{\prime}}(t_1) + a^\dag_{-\mathbf{q}^{\prime}\lambda^{\prime}} (t_1) ]   | \Psi_{\alpha}; \mathbf{q} \c1 \lambda \rangle
}
 \notag \\
P_{\alpha\beta,6} &=& \frac{(-i)^2}{2! \hbar^2} \iiint_{-\infty}^{+\infty} d t_1 d t_2 d t_3 \sum_{\mathbf{k}^{\prime}_1 \mathbf{k}^{\prime}_2 \mathbf{q}^{\prime}_1 \sigma^{\prime}_1 \sigma^{\prime}_2} \sum_{\mathbf{k}^{\prime} \mathbf{q}^{\prime} \lambda^{\prime} \sigma^{\prime}} \frac{1}{2}	U(\mathbf{k}^{\prime}_1, \mathbf{k}^{\prime}_2, \mathbf{q}^{\prime}_1; t_2 - t_3) g_{\lambda^{\prime}}(\mathbf{k}^{\prime},\mathbf{q}^{\prime})  \notag \\
&& \times 
\wick{
\langle \Psi_\beta; \mathbf{k}_{1} \c2 \sigma_1 ,\mathbf{k}_2 \c1 \sigma_2 | T_t  \c1 f^\dag_{\mathbf{k}^{\prime}_1+\mathbf{q}^{\prime}_1 \sigma^{\prime}_1}(t_2) c_{\mathbf{k}^{\prime}_1 \sigma^{\prime}_1} (t_2) \c1 d^\dag_{\mathbf{k}^{\prime}_2 -\mathbf{q}^{\prime}_1 \sigma^{\prime}_2}(t_3) c_{\mathbf{k}^{\prime}_2 \sigma^{\prime}_2 }(t_3) \c2 f^\dag_{\mathbf{k}^{\prime}+\mathbf{q}^{\prime} \sigma^{\prime}}(t_1) \c1 d_{\mathbf{k}^{\prime}\sigma^{\prime}} (t_1) [ \c1 a_{ \mathbf{q}^{\prime}\lambda^{\prime}}(t_1) + a^\dag_{-\mathbf{q}^{\prime}\lambda^{\prime}} (t_1) ]   | \Psi_{\alpha}; \mathbf{q} \c1 \lambda \rangle
}
 \notag \\
 P_{\alpha\beta,7} &=& \frac{(-i)^2}{2! \hbar^2} \iiint_{-\infty}^{+\infty} d t_1 d t_2 d t_3 \sum_{\mathbf{k}^{\prime}_1 \mathbf{k}^{\prime}_2 \mathbf{q}^{\prime}_1 \sigma^{\prime}_1 \sigma^{\prime}_2} \sum_{\mathbf{k}^{\prime} \mathbf{q}^{\prime} \lambda^{\prime} \sigma^{\prime}} \frac{1}{2}	U(\mathbf{k}^{\prime}_1, \mathbf{k}^{\prime}_2, \mathbf{q}^{\prime}_1; t_2 - t_3) g_{\lambda^{\prime}}(\mathbf{k}^{\prime},\mathbf{q}^{\prime})  \label{eqnC.1} \\
&& \times 
\wick{
\langle \Psi_\beta; \mathbf{k}_{1} \c1 \sigma_1 ,\mathbf{k}_2 \c2 \sigma_2 | T_t \c3 d^\dag_{\mathbf{k}^{\prime}_1+\mathbf{q}^{\prime}_1 \sigma^{\prime}_1}(t_2) c_{\mathbf{k}^{\prime}_1 \sigma^{\prime}_1} (t_2) \c1 f^\dag_{\mathbf{k}^{\prime}_2 -\mathbf{q}^{\prime}_1 \sigma^{\prime}_2}(t_3) c_{\mathbf{k}^{\prime}_2 \sigma^{\prime}_2 }(t_3) \c2 f^\dag_{\mathbf{k}^{\prime}+\mathbf{q}^{\prime} \sigma^{\prime}}(t_1) \c3 d_{\mathbf{k}^{\prime}\sigma^{\prime}} (t_1) [ \c1 a_{ \mathbf{q}^{\prime}\lambda^{\prime}}(t_1) + a^\dag_{-\mathbf{q}^{\prime}\lambda^{\prime}} (t_1) ]   | \Psi_{\alpha}; \mathbf{q} \c1 \lambda \rangle
}
 \notag \\
P_{\alpha\beta,8} &=& \frac{(-i)^2}{2! \hbar^2} \iiint_{-\infty}^{+\infty} d t_1 d t_2 d t_3 \sum_{\mathbf{k}^{\prime}_1 \mathbf{k}^{\prime}_2 \mathbf{q}^{\prime}_1 \sigma^{\prime}_1 \sigma^{\prime}_2} \sum_{\mathbf{k}^{\prime} \mathbf{q}^{\prime} \lambda^{\prime} \sigma^{\prime}} \frac{1}{2}	U(\mathbf{k}^{\prime}_1, \mathbf{k}^{\prime}_2, \mathbf{q}^{\prime}_1; t_2 - t_3) g_{\lambda^{\prime}}(\mathbf{k}^{\prime},\mathbf{q}^{\prime})  \notag \\
&& \times 
\wick{
\langle \Psi_\beta; \mathbf{k}_{1} \c2 \sigma_1 ,\mathbf{k}_2 \c1 \sigma_2 | T_t \c3 d^\dag_{\mathbf{k}^{\prime}_1+\mathbf{q}^{\prime}_1 \sigma^{\prime}_1}(t_2) c_{\mathbf{k}^{\prime}_1 \sigma^{\prime}_1} (t_2) \c1 f^\dag_{\mathbf{k}^{\prime}_2 -\mathbf{q}^{\prime}_1 \sigma^{\prime}_2}(t_3) c_{\mathbf{k}^{\prime}_2 \sigma^{\prime}_2 }(t_3) \c2 f^\dag_{\mathbf{k}^{\prime}+\mathbf{q}^{\prime} \sigma^{\prime}}(t_1) \c3 d_{\mathbf{k}^{\prime}\sigma^{\prime}} (t_1) [ \c1 a_{ \mathbf{q}^{\prime}\lambda^{\prime}}(t_1) + a^\dag_{-\mathbf{q}^{\prime}\lambda^{\prime}} (t_1) ]   | \Psi_{\alpha}; \mathbf{q} \c1 \lambda \rangle
}
 \notag
\end{eqnarray} 
With a similar calculation to type-I photoemission, we can obtain Eqs. (\ref{eqn2.16}), (\ref{eqn2.17}) and (\ref{eqn2.19}).

\end{widetext}




\begin{thebibliography}{34}%
\makeatletter
\providecommand \@ifxundefined [1]{%
 \@ifx{#1\undefined}
}%
\providecommand \@ifnum [1]{%
 \ifnum #1\expandafter \@firstoftwo
 \else \expandafter \@secondoftwo
 \fi
}%
\providecommand \@ifx [1]{%
 \ifx #1\expandafter \@firstoftwo
 \else \expandafter \@secondoftwo
 \fi
}%
\providecommand \natexlab [1]{#1}%
\providecommand \enquote  [1]{#1}%
\providecommand \bibnamefont  [1]{#1}%
\providecommand \bibfnamefont [1]{#1}%
\providecommand \citenamefont [1]{#1}%
\providecommand \href@noop [0]{\@secondoftwo}%
\providecommand \href [0]{\begingroup \@sanitize@url \@href}%
\providecommand \@href[1]{\@@startlink{#1}\@@href}%
\providecommand \@@href[1]{\endgroup#1\@@endlink}%
\providecommand \@sanitize@url [0]{\catcode `\\12\catcode `\$12\catcode
  `\&12\catcode `\#12\catcode `\^12\catcode `\_12\catcode `\%12\relax}%
\providecommand \@@startlink[1]{}%
\providecommand \@@endlink[0]{}%
\providecommand \url  [0]{\begingroup\@sanitize@url \@url }%
\providecommand \@url [1]{\endgroup\@href {#1}{\urlprefix }}%
\providecommand \urlprefix  [0]{URL }%
\providecommand \Eprint [0]{\href }%
\providecommand \doibase [0]{http://dx.doi.org/}%
\providecommand \selectlanguage [0]{\@gobble}%
\providecommand \bibinfo  [0]{\@secondoftwo}%
\providecommand \bibfield  [0]{\@secondoftwo}%
\providecommand \translation [1]{[#1]}%
\providecommand \BibitemOpen [0]{}%
\providecommand \bibitemStop [0]{}%
\providecommand \bibitemNoStop [0]{.\EOS\space}%
\providecommand \EOS [0]{\spacefactor3000\relax}%
\providecommand \BibitemShut  [1]{\csname bibitem#1\endcsname}%
\let\auto@bib@innerbib\@empty
\bibitem [{\citenamefont {Lee}\ \emph {et~al.}(2006)\citenamefont {Lee},
  \citenamefont {Nagaosa},\ and\ \citenamefont {Wen}}]{PALeeRMP2006}%
  \BibitemOpen
  \bibfield  {author} {\bibinfo {author} {\bibfnamefont {P.~A.}\ \bibnamefont
  {Lee}}, \bibinfo {author} {\bibfnamefont {N.}~\bibnamefont {Nagaosa}}, \ and\
  \bibinfo {author} {\bibfnamefont {X.-G.}\ \bibnamefont {Wen}},\ }\enquote
  {\bibinfo {title} {Doping a {M}ott insulator: Physics of high-temperature
  superconductivity},}\ \href {\doibase 10.1103/RevModPhys.78.17} {\bibfield
  {journal} {\bibinfo  {journal} {Rev. Mod. Phys.}\ }\textbf {\bibinfo {volume}
  {78}},\ \bibinfo {pages} {17} (\bibinfo {year} {2006})}\BibitemShut {NoStop}%
\bibitem [{\citenamefont {Stewart}(2001)}]{StewartNFLRMP2001}%
  \BibitemOpen
  \bibfield  {author} {\bibinfo {author} {\bibfnamefont {G.~R.}\ \bibnamefont
  {Stewart}},\ }\enquote {\bibinfo {title} {Non-{Fermi}-liquid behavior in $d$-
  and $f$-electron metals},}\ \href {\doibase 10.1103/RevModPhys.73.797}
  {\bibfield  {journal} {\bibinfo  {journal} {Rev. Mod. Phys.}\ }\textbf
  {\bibinfo {volume} {73}},\ \bibinfo {pages} {797} (\bibinfo {year}
  {2001})}\BibitemShut {NoStop}%
\bibitem [{\citenamefont {Fradkin}\ \emph {et~al.}(2010)\citenamefont
  {Fradkin}, \citenamefont {Kivelson}, \citenamefont {Lawler}, \citenamefont
  {Eisenstein},\ and\ \citenamefont {Mackenzie}}]{FradkinARCMP2010}%
  \BibitemOpen
  \bibfield  {author} {\bibinfo {author} {\bibfnamefont {E.}~\bibnamefont
  {Fradkin}}, \bibinfo {author} {\bibfnamefont {S.~A.}\ \bibnamefont
  {Kivelson}}, \bibinfo {author} {\bibfnamefont {M.~J.}\ \bibnamefont
  {Lawler}}, \bibinfo {author} {\bibfnamefont {J.~P.}\ \bibnamefont
  {Eisenstein}}, \ and\ \bibinfo {author} {\bibfnamefont {A.~P.}\ \bibnamefont
  {Mackenzie}},\ }\enquote {\bibinfo {title} {Nematic {F}ermi fluids in
  condensed matter physics},}\ \href {\doibase
  10.1146/annurev-conmatphys-070909-103925} {\bibfield  {journal} {\bibinfo
  {journal} {Annu. Rev. Condens. Matter Phys.}\ }\textbf {\bibinfo {volume}
  {1}},\ \bibinfo {pages} {153} (\bibinfo {year} {2010})}\BibitemShut {NoStop}%
\bibitem [{\citenamefont {Stewart}(2011)}]{StewartFeSCRMP2011}%
  \BibitemOpen
  \bibfield  {author} {\bibinfo {author} {\bibfnamefont {G.~R.}\ \bibnamefont
  {Stewart}},\ }\enquote {\bibinfo {title} {Superconductivity in iron
  compounds},}\ \href {\doibase 10.1103/RevModPhys.83.1589} {\bibfield
  {journal} {\bibinfo  {journal} {Rev. Mod. Phys.}\ }\textbf {\bibinfo {volume}
  {83}},\ \bibinfo {pages} {1589} (\bibinfo {year} {2011})}\BibitemShut
  {NoStop}%
\bibitem [{\citenamefont {Chen}\ \emph {et~al.}(2014)\citenamefont {Chen},
  \citenamefont {Dai}, \citenamefont {Feng}, \citenamefont {Xiang},\ and\
  \citenamefont {Zhang}}]{ChenXHNAR2014}%
  \BibitemOpen
  \bibfield  {author} {\bibinfo {author} {\bibfnamefont {X.}~\bibnamefont
  {Chen}}, \bibinfo {author} {\bibfnamefont {P.}~\bibnamefont {Dai}}, \bibinfo
  {author} {\bibfnamefont {D.}~\bibnamefont {Feng}}, \bibinfo {author}
  {\bibfnamefont {T.}~\bibnamefont {Xiang}}, \ and\ \bibinfo {author}
  {\bibfnamefont {F.-C.}\ \bibnamefont {Zhang}},\ }\enquote {\bibinfo {title}
  {{Iron-based high transition temperature superconductors}},}\ \href {\doibase
  10.1093/nsr/nwu007} {\bibfield  {journal} {\bibinfo  {journal} {Natl. Sci.
  Rev.}\ }\textbf {\bibinfo {volume} {1}},\ \bibinfo {pages} {371} (\bibinfo
  {year} {2014})}\BibitemShut {NoStop}%
\bibitem [{\citenamefont {Su}\ and\ \citenamefont {Zhang}(2020)}]{SuZhang2020}%
  \BibitemOpen
  \bibfield  {author} {\bibinfo {author} {\bibfnamefont {Y.}~\bibnamefont
  {Su}}\ and\ \bibinfo {author} {\bibfnamefont {C.}~\bibnamefont {Zhang}},\
  }\enquote {\bibinfo {title} {Coincidence angle-resolved photoemission
  spectroscopy: Proposal for detection of two-particle correlations},}\ \href
  {\doibase 10.1103/PhysRevB.101.205110} {\bibfield  {journal} {\bibinfo
  {journal} {Phys. Rev. B}\ }\textbf {\bibinfo {volume} {101}},\ \bibinfo
  {pages} {205110} (\bibinfo {year} {2020})}\BibitemShut {NoStop}%
\bibitem [{\citenamefont {Devereaux}\ \emph {et~al.}(2023)\citenamefont
  {Devereaux}, \citenamefont {Claassen}, \citenamefont {Huang}, \citenamefont
  {Zaletel}, \citenamefont {Moore}, \citenamefont {Morr}, \citenamefont
  {Mahmood}, \citenamefont {Abbamonte},\ and\ \citenamefont
  {Shen}}]{DevereauxPRB2023}%
  \BibitemOpen
  \bibfield  {author} {\bibinfo {author} {\bibfnamefont {T.~P.}\ \bibnamefont
  {Devereaux}}, \bibinfo {author} {\bibfnamefont {M.}~\bibnamefont {Claassen}},
  \bibinfo {author} {\bibfnamefont {X.-X.}\ \bibnamefont {Huang}}, \bibinfo
  {author} {\bibfnamefont {M.}~\bibnamefont {Zaletel}}, \bibinfo {author}
  {\bibfnamefont {J.~E.}\ \bibnamefont {Moore}}, \bibinfo {author}
  {\bibfnamefont {D.}~\bibnamefont {Morr}}, \bibinfo {author} {\bibfnamefont
  {F.}~\bibnamefont {Mahmood}}, \bibinfo {author} {\bibfnamefont
  {P.}~\bibnamefont {Abbamonte}}, \ and\ \bibinfo {author} {\bibfnamefont
  {Z.-X.}\ \bibnamefont {Shen}},\ }\enquote {\bibinfo {title} {Angle-resolved
  pair photoemission theory for correlated electrons},}\ \href {\doibase
  10.1103/PhysRevB.108.165134} {\bibfield  {journal} {\bibinfo  {journal}
  {Phys. Rev. B}\ }\textbf {\bibinfo {volume} {108}},\ \bibinfo {pages}
  {165134} (\bibinfo {year} {2023})}\BibitemShut {NoStop}%
\bibitem [{\citenamefont {Cao}\ and\ \citenamefont {Su}(2024)}]{CaoSu2024-2}%
  \BibitemOpen
  \bibfield  {author} {\bibinfo {author} {\bibfnamefont {D.}~\bibnamefont
  {Cao}}\ and\ \bibinfo {author} {\bibfnamefont {Y.}~\bibnamefont {Su}},\
  }\href@noop {} {\enquote {\bibinfo {title} {Post-experiment coincidence
  detection techniques for direct detection of two-body correlations},}\ }
  (\bibinfo {year} {2024}),\ \Eprint {http://arxiv.org/abs/2308.16746}
  {arXiv:2308.16746 [cond-mat.supr-con]} \BibitemShut {NoStop}%
\bibitem [{\citenamefont {Su}\ \emph {et~al.}(2021)\citenamefont {Su},
  \citenamefont {Wang},\ and\ \citenamefont {Zhang}}]{SucINS2021}%
  \BibitemOpen
  \bibfield  {author} {\bibinfo {author} {\bibfnamefont {Y.}~\bibnamefont
  {Su}}, \bibinfo {author} {\bibfnamefont {S.}~\bibnamefont {Wang}}, \ and\
  \bibinfo {author} {\bibfnamefont {C.}~\bibnamefont {Zhang}},\ }\enquote
  {\bibinfo {title} {Coincidence inelastic neutron scattering for detection of
  two-spin magnetic correlations},}\ \href {\doibase
  10.1103/PhysRevB.103.054431} {\bibfield  {journal} {\bibinfo  {journal}
  {Phys. Rev. B}\ }\textbf {\bibinfo {volume} {103}},\ \bibinfo {pages}
  {054431} (\bibinfo {year} {2021})}\BibitemShut {NoStop}%
\bibitem [{\citenamefont {Haak}\ \emph {et~al.}(1978)\citenamefont {Haak},
  \citenamefont {Sawatzky},\ and\ \citenamefont {Thomas}}]{HaakAugerPRL1978}%
  \BibitemOpen
  \bibfield  {author} {\bibinfo {author} {\bibfnamefont {H.~W.}\ \bibnamefont
  {Haak}}, \bibinfo {author} {\bibfnamefont {G.~A.}\ \bibnamefont {Sawatzky}},
  \ and\ \bibinfo {author} {\bibfnamefont {T.~D.}\ \bibnamefont {Thomas}},\
  }\enquote {\bibinfo {title} {Auger-photoelectron coincidence measurements in
  copper},}\ \href {\doibase 10.1103/PhysRevLett.41.1825} {\bibfield  {journal}
  {\bibinfo  {journal} {Phys. Rev. Lett.}\ }\textbf {\bibinfo {volume} {41}},\
  \bibinfo {pages} {1825} (\bibinfo {year} {1978})}\BibitemShut {NoStop}%
\bibitem [{\citenamefont {Weber}\ \emph {et~al.}(2005)\citenamefont {Weber},
  \citenamefont {Giessen}, \citenamefont {Weckenbrock}, \citenamefont
  {Urbasch}, \citenamefont {Staudte}, \citenamefont {Spielberger},
  \citenamefont {Jagutzki}, \citenamefont {Mergel}, \citenamefont {Vollmer},\
  and\ \citenamefont {D\"orner}}]{WeberNature2005}%
  \BibitemOpen
  \bibfield  {author} {\bibinfo {author} {\bibfnamefont {T.}~\bibnamefont
  {Weber}}, \bibinfo {author} {\bibfnamefont {H.}~\bibnamefont {Giessen}},
  \bibinfo {author} {\bibfnamefont {M.}~\bibnamefont {Weckenbrock}}, \bibinfo
  {author} {\bibfnamefont {G.}~\bibnamefont {Urbasch}}, \bibinfo {author}
  {\bibfnamefont {A.}~\bibnamefont {Staudte}}, \bibinfo {author} {\bibfnamefont
  {L.}~\bibnamefont {Spielberger}}, \bibinfo {author} {\bibfnamefont
  {O.}~\bibnamefont {Jagutzki}}, \bibinfo {author} {\bibfnamefont
  {V.}~\bibnamefont {Mergel}}, \bibinfo {author} {\bibfnamefont
  {M.}~\bibnamefont {Vollmer}}, \ and\ \bibinfo {author} {\bibfnamefont
  {R.}~\bibnamefont {D\"orner}},\ }\enquote {\bibinfo {title} {Correlated
  electron emission in multiphoton double ionization},}\ \href {\doibase
  10.1038/35015033} {\bibfield  {journal} {\bibinfo  {journal} {Nature
  (London)}\ }\textbf {\bibinfo {volume} {405}},\ \bibinfo {pages} {658}
  (\bibinfo {year} {2005})}\BibitemShut {NoStop}%
\bibitem [{\citenamefont {Vanroose}\ \emph {et~al.}(2005)\citenamefont
  {Vanroose}, \citenamefont {Martin}, \citenamefont {Rescigno},\ and\
  \citenamefont {William~McCurdy}}]{2005H2Science}%
  \BibitemOpen
  \bibfield  {author} {\bibinfo {author} {\bibfnamefont {W.}~\bibnamefont
  {Vanroose}}, \bibinfo {author} {\bibfnamefont {F.}~\bibnamefont {Martin}},
  \bibinfo {author} {\bibfnamefont {T.~N.}\ \bibnamefont {Rescigno}}, \ and\
  \bibinfo {author} {\bibfnamefont {C.}~\bibnamefont {William~McCurdy}},\
  }\enquote {\bibinfo {title} {Complete photo-induced breakup of the {H}$_2$
  molecule as a probe of molecular electron correlation},}\ \href {\doibase
  10.1126/science.1120263} {\bibfield  {journal} {\bibinfo  {journal}
  {Science}\ }\textbf {\bibinfo {volume} {310}},\ \bibinfo {pages} {1787}
  (\bibinfo {year} {2005})}\BibitemShut {NoStop}%
\bibitem [{\citenamefont {Akoury}\ \emph {et~al.}(2007)\citenamefont {Akoury},
  \citenamefont {Kreidi}, \citenamefont {Jahnke}, \citenamefont {Weber},
  \citenamefont {Staudte}, \citenamefont {Sch\"offler}, \citenamefont
  {Neumann}, \citenamefont {Titze}, \citenamefont {Schmidt}, \citenamefont
  {Czasch}, \citenamefont {Jagutzki}, \citenamefont {Costa~Fraga},
  \citenamefont {Grisenti}, \citenamefont {Diez~Muino}, \citenamefont
  {Cherepkov}, \citenamefont {Semenov}, \citenamefont {Ranitovic},
  \citenamefont {Cocke}, \citenamefont {Osipov}, \citenamefont {Adaniya},
  \citenamefont {Thompson}, \citenamefont {Prior}, \citenamefont {Belkacem},
  \citenamefont {Landers}, \citenamefont {Schmidt-B\"ocking},\ and\
  \citenamefont {D\"orner}}]{2007H2Science}%
  \BibitemOpen
  \bibfield  {author} {\bibinfo {author} {\bibfnamefont {D.}~\bibnamefont
  {Akoury}}, \bibinfo {author} {\bibfnamefont {K.}~\bibnamefont {Kreidi}},
  \bibinfo {author} {\bibfnamefont {T.}~\bibnamefont {Jahnke}}, \bibinfo
  {author} {\bibfnamefont {T.}~\bibnamefont {Weber}}, \bibinfo {author}
  {\bibfnamefont {A.}~\bibnamefont {Staudte}}, \bibinfo {author} {\bibfnamefont
  {M.}~\bibnamefont {Sch\"offler}}, \bibinfo {author} {\bibfnamefont
  {N.}~\bibnamefont {Neumann}}, \bibinfo {author} {\bibfnamefont
  {J.}~\bibnamefont {Titze}}, \bibinfo {author} {\bibfnamefont {L.~P.~H.}\
  \bibnamefont {Schmidt}}, \bibinfo {author} {\bibfnamefont {A.}~\bibnamefont
  {Czasch}}, \bibinfo {author} {\bibfnamefont {O.}~\bibnamefont {Jagutzki}},
  \bibinfo {author} {\bibfnamefont {R.~A.}\ \bibnamefont {Costa~Fraga}},
  \bibinfo {author} {\bibfnamefont {R.~E.}\ \bibnamefont {Grisenti}}, \bibinfo
  {author} {\bibfnamefont {R.}~\bibnamefont {Diez~Muino}}, \bibinfo {author}
  {\bibfnamefont {N.~A.}\ \bibnamefont {Cherepkov}}, \bibinfo {author}
  {\bibfnamefont {S.~K.}\ \bibnamefont {Semenov}}, \bibinfo {author}
  {\bibfnamefont {P.}~\bibnamefont {Ranitovic}}, \bibinfo {author}
  {\bibfnamefont {C.~L.}\ \bibnamefont {Cocke}}, \bibinfo {author}
  {\bibfnamefont {T.}~\bibnamefont {Osipov}}, \bibinfo {author} {\bibfnamefont
  {H.}~\bibnamefont {Adaniya}}, \bibinfo {author} {\bibfnamefont {J.~C.}\
  \bibnamefont {Thompson}}, \bibinfo {author} {\bibfnamefont {M.~H.}\
  \bibnamefont {Prior}}, \bibinfo {author} {\bibfnamefont {A.}~\bibnamefont
  {Belkacem}}, \bibinfo {author} {\bibfnamefont {A.~L.}\ \bibnamefont
  {Landers}}, \bibinfo {author} {\bibfnamefont {H.}~\bibnamefont
  {Schmidt-B\"ocking}}, \ and\ \bibinfo {author} {\bibfnamefont
  {R.}~\bibnamefont {D\"orner}},\ }\enquote {\bibinfo {title} {The simplest
  double slit: Interference and entanglement in double photoionization of
  {H}$_2$},}\ \href {\doibase 10.1126/science.1144959} {\bibfield  {journal}
  {\bibinfo  {journal} {Science}\ }\textbf {\bibinfo {volume} {318}},\ \bibinfo
  {pages} {949} (\bibinfo {year} {2007})}\BibitemShut {NoStop}%
\bibitem [{\citenamefont {Fominykh}\ \emph {et~al.}(2002)\citenamefont
  {Fominykh}, \citenamefont {Berakdar}, \citenamefont {Henk},\ and\
  \citenamefont {Bruno}}]{FominykhPRL2002}%
  \BibitemOpen
  \bibfield  {author} {\bibinfo {author} {\bibfnamefont {N.}~\bibnamefont
  {Fominykh}}, \bibinfo {author} {\bibfnamefont {J.}~\bibnamefont {Berakdar}},
  \bibinfo {author} {\bibfnamefont {J.}~\bibnamefont {Henk}}, \ and\ \bibinfo
  {author} {\bibfnamefont {P.}~\bibnamefont {Bruno}},\ }\enquote {\bibinfo
  {title} {Spectroscopy of the electron-electron interaction in solids},}\
  \href {\doibase 10.1103/PhysRevLett.89.086402} {\bibfield  {journal}
  {\bibinfo  {journal} {Phys. Rev. Lett.}\ }\textbf {\bibinfo {volume} {89}},\
  \bibinfo {pages} {086402} (\bibinfo {year} {2002})}\BibitemShut {NoStop}%
\bibitem [{\citenamefont {Tr\"utzschler}\ \emph {et~al.}(2017)\citenamefont
  {Tr\"utzschler}, \citenamefont {Huth}, \citenamefont {Chiang}, \citenamefont
  {Kamrla}, \citenamefont {Schumann}, \citenamefont {Kirschner},\ and\
  \citenamefont {Widdra}}]{ChiangPRL2017}%
  \BibitemOpen
  \bibfield  {author} {\bibinfo {author} {\bibfnamefont {A.}~\bibnamefont
  {Tr\"utzschler}}, \bibinfo {author} {\bibfnamefont {M.}~\bibnamefont {Huth}},
  \bibinfo {author} {\bibfnamefont {C.-T.}\ \bibnamefont {Chiang}}, \bibinfo
  {author} {\bibfnamefont {R.}~\bibnamefont {Kamrla}}, \bibinfo {author}
  {\bibfnamefont {F.~O.}\ \bibnamefont {Schumann}}, \bibinfo {author}
  {\bibfnamefont {J.}~\bibnamefont {Kirschner}}, \ and\ \bibinfo {author}
  {\bibfnamefont {W.}~\bibnamefont {Widdra}},\ }\enquote {\bibinfo {title}
  {Band-resolved double photoemission spectroscopy on correlated valence
  electron pairs in metals},}\ \href {\doibase 10.1103/PhysRevLett.118.136401}
  {\bibfield  {journal} {\bibinfo  {journal} {Phys. Rev. Lett.}\ }\textbf
  {\bibinfo {volume} {118}},\ \bibinfo {pages} {136401} (\bibinfo {year}
  {2017})}\BibitemShut {NoStop}%
\bibitem [{\citenamefont {Aliaev}\ \emph {et~al.}(2018)\citenamefont {Aliaev},
  \citenamefont {Kostanovskiy}, \citenamefont {Kirschner},\ and\ \citenamefont
  {Schumann}}]{Aliaev2018167}%
  \BibitemOpen
  \bibfield  {author} {\bibinfo {author} {\bibfnamefont {Y.}~\bibnamefont
  {Aliaev}}, \bibinfo {author} {\bibfnamefont {I.}~\bibnamefont
  {Kostanovskiy}}, \bibinfo {author} {\bibfnamefont {J.}~\bibnamefont
  {Kirschner}}, \ and\ \bibinfo {author} {\bibfnamefont {F.}~\bibnamefont
  {Schumann}},\ }\enquote {\bibinfo {title} {Electron pair emission from a {P}b
  surface at room temperature},}\ \href {\doibase
  https://doi.org/10.1016/j.susc.2018.06.013} {\bibfield  {journal} {\bibinfo
  {journal} {Surf. Sci.}\ }\textbf {\bibinfo {volume} {677}},\ \bibinfo {pages}
  {167} (\bibinfo {year} {2018})}\BibitemShut {NoStop}%
\bibitem [{\citenamefont {Kouzakov}\ and\ \citenamefont
  {Berakdar}(2003)}]{BerakdarSCPRL2003}%
  \BibitemOpen
  \bibfield  {author} {\bibinfo {author} {\bibfnamefont {K.~A.}\ \bibnamefont
  {Kouzakov}}\ and\ \bibinfo {author} {\bibfnamefont {J.}~\bibnamefont
  {Berakdar}},\ }\enquote {\bibinfo {title} {Photoinduced emission of {C}ooper
  pairs from superconductors},}\ \href {\doibase 10.1103/PhysRevLett.91.257007}
  {\bibfield  {journal} {\bibinfo  {journal} {Phys. Rev. Lett.}\ }\textbf
  {\bibinfo {volume} {91}},\ \bibinfo {pages} {257007} (\bibinfo {year}
  {2003})}\BibitemShut {NoStop}%
\bibitem [{\citenamefont {Berakdar}(1998)}]{BerakdarPRB1998}%
  \BibitemOpen
  \bibfield  {author} {\bibinfo {author} {\bibfnamefont {J.}~\bibnamefont
  {Berakdar}},\ }\enquote {\bibinfo {title} {Emission of correlated electron
  pairs following single-photon absorption by solids and surfaces},}\ \href
  {\doibase 10.1103/PhysRevB.58.9808} {\bibfield  {journal} {\bibinfo
  {journal} {Phys. Rev. B}\ }\textbf {\bibinfo {volume} {58}},\ \bibinfo
  {pages} {9808} (\bibinfo {year} {1998})}\BibitemShut {NoStop}%
\bibitem [{\citenamefont {Fominykh}\ \emph {et~al.}(2000)\citenamefont
  {Fominykh}, \citenamefont {Henk}, \citenamefont {Berakdar}, \citenamefont
  {Bruno}, \citenamefont {Gollisch},\ and\ \citenamefont
  {Feder}}]{Fominykh2000665}%
  \BibitemOpen
  \bibfield  {author} {\bibinfo {author} {\bibfnamefont {N.}~\bibnamefont
  {Fominykh}}, \bibinfo {author} {\bibfnamefont {J.}~\bibnamefont {Henk}},
  \bibinfo {author} {\bibfnamefont {J.}~\bibnamefont {Berakdar}}, \bibinfo
  {author} {\bibfnamefont {P.}~\bibnamefont {Bruno}}, \bibinfo {author}
  {\bibfnamefont {H.}~\bibnamefont {Gollisch}}, \ and\ \bibinfo {author}
  {\bibfnamefont {R.}~\bibnamefont {Feder}},\ }\enquote {\bibinfo {title}
  {Theory of two-electron photoemission from surfaces},}\ \href {\doibase
  https://doi.org/10.1016/S0038-1098(99)00568-2} {\bibfield  {journal}
  {\bibinfo  {journal} {Solid State Commun.}\ }\textbf {\bibinfo {volume}
  {113}},\ \bibinfo {pages} {665} (\bibinfo {year} {2000})}\BibitemShut
  {NoStop}%
\bibitem [{\citenamefont {Pavlyukh}\ \emph {et~al.}(2015)\citenamefont
  {Pavlyukh}, \citenamefont {Sch\"uler},\ and\ \citenamefont
  {Berakdar}}]{BerakdarPRB2015}%
  \BibitemOpen
  \bibfield  {author} {\bibinfo {author} {\bibfnamefont {Y.}~\bibnamefont
  {Pavlyukh}}, \bibinfo {author} {\bibfnamefont {M.}~\bibnamefont {Sch\"uler}},
  \ and\ \bibinfo {author} {\bibfnamefont {J.}~\bibnamefont {Berakdar}},\
  }\enquote {\bibinfo {title} {Single- or double-electron emission within the
  {K}eldysh nonequilibrium {G}reen's function and {F}eshbach projection
  operator techniques},}\ \href {\doibase 10.1103/PhysRevB.91.155116}
  {\bibfield  {journal} {\bibinfo  {journal} {Phys. Rev. B}\ }\textbf {\bibinfo
  {volume} {91}},\ \bibinfo {pages} {155116} (\bibinfo {year}
  {2015})}\BibitemShut {NoStop}%
\bibitem [{\citenamefont {Damascelli}\ \emph {et~al.}(2003)\citenamefont
  {Damascelli}, \citenamefont {Hussain},\ and\ \citenamefont
  {Shen}}]{ShenRMP2003}%
  \BibitemOpen
  \bibfield  {author} {\bibinfo {author} {\bibfnamefont {A.}~\bibnamefont
  {Damascelli}}, \bibinfo {author} {\bibfnamefont {Z.}~\bibnamefont {Hussain}},
  \ and\ \bibinfo {author} {\bibfnamefont {Z.-X.}\ \bibnamefont {Shen}},\
  }\enquote {\bibinfo {title} {Angle-resolved photoemission studies of the
  cuprate superconductors},}\ \href {\doibase 10.1103/RevModPhys.75.473}
  {\bibfield  {journal} {\bibinfo  {journal} {Rev. Mod. Phys.}\ }\textbf
  {\bibinfo {volume} {75}},\ \bibinfo {pages} {473} (\bibinfo {year}
  {2003})}\BibitemShut {NoStop}%
\bibitem [{\citenamefont {Freericks}\ \emph {et~al.}(2009)\citenamefont
  {Freericks}, \citenamefont {Krishnamurthy},\ and\ \citenamefont
  {Pruschke}}]{FreericksTRARPESPRL2009}%
  \BibitemOpen
  \bibfield  {author} {\bibinfo {author} {\bibfnamefont {J.~K.}\ \bibnamefont
  {Freericks}}, \bibinfo {author} {\bibfnamefont {H.~R.}\ \bibnamefont
  {Krishnamurthy}}, \ and\ \bibinfo {author} {\bibfnamefont {T.}~\bibnamefont
  {Pruschke}},\ }\enquote {\bibinfo {title} {Theoretical description of
  time-resolved photoemission spectroscopy: Application to pump-probe
  experiments},}\ \href {\doibase 10.1103/PhysRevLett.102.136401} {\bibfield
  {journal} {\bibinfo  {journal} {Phys. Rev. Lett.}\ }\textbf {\bibinfo
  {volume} {102}},\ \bibinfo {pages} {136401} (\bibinfo {year}
  {2009})}\BibitemShut {NoStop}%
\bibitem [{\citenamefont {Gao}\ \emph {et~al.}(2020)\citenamefont {Gao},
  \citenamefont {Schlawin}, \citenamefont {Buzzi}, \citenamefont {Cavalleri},\
  and\ \citenamefont {Jaksch}}]{JakschPairingCavityPRL2020}%
  \BibitemOpen
  \bibfield  {author} {\bibinfo {author} {\bibfnamefont {H.}~\bibnamefont
  {Gao}}, \bibinfo {author} {\bibfnamefont {F.}~\bibnamefont {Schlawin}},
  \bibinfo {author} {\bibfnamefont {M.}~\bibnamefont {Buzzi}}, \bibinfo
  {author} {\bibfnamefont {A.}~\bibnamefont {Cavalleri}}, \ and\ \bibinfo
  {author} {\bibfnamefont {D.}~\bibnamefont {Jaksch}},\ }\enquote {\bibinfo
  {title} {Photoinduced electron pairing in a driven cavity},}\ \href {\doibase
  10.1103/PhysRevLett.125.053602} {\bibfield  {journal} {\bibinfo  {journal}
  {Phys. Rev. Lett.}\ }\textbf {\bibinfo {volume} {125}},\ \bibinfo {pages}
  {053602} (\bibinfo {year} {2020})}\BibitemShut {NoStop}%
\bibitem [{\citenamefont {Salpeter}\ and\ \citenamefont
  {Bethe}(1951)}]{SalpeterBethe1951}%
  \BibitemOpen
  \bibfield  {author} {\bibinfo {author} {\bibfnamefont {E.~E.}\ \bibnamefont
  {Salpeter}}\ and\ \bibinfo {author} {\bibfnamefont {H.~A.}\ \bibnamefont
  {Bethe}},\ }\enquote {\bibinfo {title} {A relativistic equation for
  bound-state problems},}\ \href {\doibase 10.1103/PhysRev.84.1232} {\bibfield
  {journal} {\bibinfo  {journal} {Phys. Rev.}\ }\textbf {\bibinfo {volume}
  {84}},\ \bibinfo {pages} {1232} (\bibinfo {year} {1951})}\BibitemShut
  {NoStop}%
\bibitem [{\citenamefont {Gell-Mann}\ and\ \citenamefont
  {Low}(1951)}]{GellmanLowBS1951}%
  \BibitemOpen
  \bibfield  {author} {\bibinfo {author} {\bibfnamefont {M.}~\bibnamefont
  {Gell-Mann}}\ and\ \bibinfo {author} {\bibfnamefont {F.}~\bibnamefont
  {Low}},\ }\enquote {\bibinfo {title} {Bound states in quantum field
  theory},}\ \href {\doibase 10.1103/PhysRev.84.350} {\bibfield  {journal}
  {\bibinfo  {journal} {Phys. Rev.}\ }\textbf {\bibinfo {volume} {84}},\
  \bibinfo {pages} {350} (\bibinfo {year} {1951})}\BibitemShut {NoStop}%
\bibitem [{\citenamefont {Mahmood}\ \emph {et~al.}(2022)\citenamefont
  {Mahmood}, \citenamefont {Devereaux}, \citenamefont {Abbamonte},\ and\
  \citenamefont {Morr}}]{MorrPRB2022}%
  \BibitemOpen
  \bibfield  {author} {\bibinfo {author} {\bibfnamefont {F.}~\bibnamefont
  {Mahmood}}, \bibinfo {author} {\bibfnamefont {T.}~\bibnamefont {Devereaux}},
  \bibinfo {author} {\bibfnamefont {P.}~\bibnamefont {Abbamonte}}, \ and\
  \bibinfo {author} {\bibfnamefont {D.~K.}\ \bibnamefont {Morr}},\ }\enquote
  {\bibinfo {title} {Distinguishing finite-momentum superconducting pairing
  states with two-electron photoemission spectroscopy},}\ \href {\doibase
  10.1103/PhysRevB.105.064515} {\bibfield  {journal} {\bibinfo  {journal}
  {Phys. Rev. B}\ }\textbf {\bibinfo {volume} {105}},\ \bibinfo {pages}
  {064515} (\bibinfo {year} {2022})}\BibitemShut {NoStop}%
\bibitem [{\citenamefont {Anderson}(1958)}]{AndersonPlasmon1958}%
  \BibitemOpen
  \bibfield  {author} {\bibinfo {author} {\bibfnamefont {P.~W.}\ \bibnamefont
  {Anderson}},\ }\enquote {\bibinfo {title} {Random-phase approximation in the
  theory of superconductivity},}\ \href {\doibase 10.1103/PhysRev.112.1900}
  {\bibfield  {journal} {\bibinfo  {journal} {Phys. Rev.}\ }\textbf {\bibinfo
  {volume} {112}},\ \bibinfo {pages} {1900} (\bibinfo {year}
  {1958})}\BibitemShut {NoStop}%
\bibitem [{\citenamefont {Negele}\ and\ \citenamefont
  {Orland}(1987)}]{NegeleOrland}%
  \BibitemOpen
  \bibfield  {author} {\bibinfo {author} {\bibfnamefont {J.~W.}\ \bibnamefont
  {Negele}}\ and\ \bibinfo {author} {\bibfnamefont {H.}~\bibnamefont
  {Orland}},\ }\href@noop {} {\emph {\bibinfo {title} {Quantum many-particle
  system}}}\ (\bibinfo  {publisher} {Addison-Wesley Publishing Company},\
  \bibinfo {year} {1987})\BibitemShut {NoStop}%
\bibitem [{\citenamefont {Kresin}\ and\ \citenamefont
  {Morawitz}(1988)}]{PlasmonsPRB1988}%
  \BibitemOpen
  \bibfield  {author} {\bibinfo {author} {\bibfnamefont {V.~Z.}\ \bibnamefont
  {Kresin}}\ and\ \bibinfo {author} {\bibfnamefont {H.}~\bibnamefont
  {Morawitz}},\ }\enquote {\bibinfo {title} {Layer plasmons and high-${T}_{c}$
  superconductivity},}\ \href {\doibase 10.1103/PhysRevB.37.7854} {\bibfield
  {journal} {\bibinfo  {journal} {Phys. Rev. B}\ }\textbf {\bibinfo {volume}
  {37}},\ \bibinfo {pages} {7854} (\bibinfo {year} {1988})}\BibitemShut
  {NoStop}%
\bibitem [{\citenamefont {Hepting}\ \emph {et~al.}(2018)\citenamefont
  {Hepting}, \citenamefont {Chaix}, \citenamefont {Huang}, \citenamefont
  {Fumagalli}, \citenamefont {Peng}, \citenamefont {Moritz}, \citenamefont
  {Kummer}, \citenamefont {Brookes}, \citenamefont {Lee}, \citenamefont
  {Hashimoto}, \citenamefont {Sarkar}, \citenamefont {He}, \citenamefont
  {Rotundu}, \citenamefont {Lee}, \citenamefont {Greene}, \citenamefont
  {Braicovich}, \citenamefont {Ghiringhelli}, \citenamefont {Shen},
  \citenamefont {Devereaux},\ and\ \citenamefont {Lee}}]{PlasmonsNature2018}%
  \BibitemOpen
  \bibfield  {author} {\bibinfo {author} {\bibfnamefont {M.}~\bibnamefont
  {Hepting}}, \bibinfo {author} {\bibfnamefont {L.}~\bibnamefont {Chaix}},
  \bibinfo {author} {\bibfnamefont {E.~W.}\ \bibnamefont {Huang}}, \bibinfo
  {author} {\bibfnamefont {R.}~\bibnamefont {Fumagalli}}, \bibinfo {author}
  {\bibfnamefont {Y.~Y.}\ \bibnamefont {Peng}}, \bibinfo {author}
  {\bibfnamefont {B.}~\bibnamefont {Moritz}}, \bibinfo {author} {\bibfnamefont
  {K.}~\bibnamefont {Kummer}}, \bibinfo {author} {\bibfnamefont {N.~B.}\
  \bibnamefont {Brookes}}, \bibinfo {author} {\bibfnamefont {W.~C.}\
  \bibnamefont {Lee}}, \bibinfo {author} {\bibfnamefont {M.}~\bibnamefont
  {Hashimoto}}, \bibinfo {author} {\bibfnamefont {T.}~\bibnamefont {Sarkar}},
  \bibinfo {author} {\bibfnamefont {J.-F.}\ \bibnamefont {He}}, \bibinfo
  {author} {\bibfnamefont {C.~R.}\ \bibnamefont {Rotundu}}, \bibinfo {author}
  {\bibfnamefont {Y.~S.}\ \bibnamefont {Lee}}, \bibinfo {author} {\bibfnamefont
  {R.~L.}\ \bibnamefont {Greene}}, \bibinfo {author} {\bibfnamefont
  {L.}~\bibnamefont {Braicovich}}, \bibinfo {author} {\bibfnamefont
  {G.}~\bibnamefont {Ghiringhelli}}, \bibinfo {author} {\bibfnamefont {Z.~X.}\
  \bibnamefont {Shen}}, \bibinfo {author} {\bibfnamefont {T.~P.}\ \bibnamefont
  {Devereaux}}, \ and\ \bibinfo {author} {\bibfnamefont {W.~S.}\ \bibnamefont
  {Lee}},\ }\enquote {\bibinfo {title} {Three-dimensional collective charge
  excitations in electron-doped copper oxide superconductors},}\ \href
  {\doibase 10.1038/s41586-018-0648-3} {\bibfield  {journal} {\bibinfo
  {journal} {Nature (London)}\ }\textbf {\bibinfo {volume} {563}},\ \bibinfo
  {pages} {374} (\bibinfo {year} {2018})}\BibitemShut {NoStop}%
\bibitem [{\citenamefont {Littlewood}\ and\ \citenamefont
  {Varma}(1981)}]{VarmaHiggsPRL1981}%
  \BibitemOpen
  \bibfield  {author} {\bibinfo {author} {\bibfnamefont {P.~B.}\ \bibnamefont
  {Littlewood}}\ and\ \bibinfo {author} {\bibfnamefont {C.~M.}\ \bibnamefont
  {Varma}},\ }\enquote {\bibinfo {title} {Gauge-invariant theory of the
  dynamical interaction of charge density waves and superconductivity},}\ \href
  {\doibase 10.1103/PhysRevLett.47.811} {\bibfield  {journal} {\bibinfo
  {journal} {Phys. Rev. Lett.}\ }\textbf {\bibinfo {volume} {47}},\ \bibinfo
  {pages} {811} (\bibinfo {year} {1981})}\BibitemShut {NoStop}%
\bibitem [{\citenamefont {Littlewood}\ and\ \citenamefont
  {Varma}(1982)}]{VarmaHiggsPRB1982}%
  \BibitemOpen
  \bibfield  {author} {\bibinfo {author} {\bibfnamefont {P.~B.}\ \bibnamefont
  {Littlewood}}\ and\ \bibinfo {author} {\bibfnamefont {C.~M.}\ \bibnamefont
  {Varma}},\ }\enquote {\bibinfo {title} {Amplitude collective modes in
  superconductors and their coupling to charge-density waves},}\ \href
  {\doibase 10.1103/PhysRevB.26.4883} {\bibfield  {journal} {\bibinfo
  {journal} {Phys. Rev. B}\ }\textbf {\bibinfo {volume} {26}},\ \bibinfo
  {pages} {4883} (\bibinfo {year} {1982})}\BibitemShut {NoStop}%
\bibitem [{\citenamefont {Leggett}(1966)}]{LeggettMode1966}%
  \BibitemOpen
  \bibfield  {author} {\bibinfo {author} {\bibfnamefont {A.~J.}\ \bibnamefont
  {Leggett}},\ }\enquote {\bibinfo {title} {{Number-phase fluctuations in
  two-band superconductors}},}\ \href {\doibase 10.1143/PTP.36.901} {\bibfield
  {journal} {\bibinfo  {journal} {Progress of Theoretical Physics}\ }\textbf
  {\bibinfo {volume} {36}},\ \bibinfo {pages} {901} (\bibinfo {year}
  {1966})}\BibitemShut {NoStop}%
\bibitem [{\citenamefont {Bruus}\ and\ \citenamefont
  {Flensberg}(2002)}]{Bruus2002}%
  \BibitemOpen
  \bibfield  {author} {\bibinfo {author} {\bibfnamefont {H.}~\bibnamefont
  {Bruus}}\ and\ \bibinfo {author} {\bibfnamefont {K.}~\bibnamefont
  {Flensberg}},\ }\href@noop {} {\emph {\bibinfo {title} {Introduction to
  many-body quantum theory in condensed matter physics}}},\ Oxford Graduate
  Texts\ (\bibinfo  {publisher} {Oxford University Press, USA},\ \bibinfo
  {year} {2002})\BibitemShut {NoStop}%
\end{thebibliography}

%

\end{document}